\documentclass[aps,reprint,superscriptaddress,prc]{revtex4-2}
\usepackage{graphicx}  
\usepackage{amsmath}   
\usepackage{amssymb}   
\usepackage{hyperref}  
\usepackage{siunitx}
\newcommand{\um}[1]{\SI{#1}{\micro\meter}} 
\newcommand{\uw}[1]{\SI{#1}{\micro\watt}}  

\begin{document}

\title{Strong nanophotonic quantum squeezing exceeding 3.5 dB in a foundry-compatible Kerr microresonator}

\author{Yichen Shen}
\affiliation{Department of Mechanical Engineering, and Institute for Physical Science \& Technology, University of Maryland, College Park, Maryland 20742, USA}

\author{Ping-Yen Hsieh}
\affiliation{Department of Mechanical Engineering, and Institute for Physical Science \& Technology, University of Maryland, College Park, Maryland 20742, USA}
\affiliation{Department of Photonics, College of Electrical and Computer Engineering, National Yang Ming Chiao Tung University, Hsinchu 30069, Taiwan}

\author{Sashank Kaushik Sridhar}
\affiliation{Department of Mechanical Engineering, and Institute for Physical Science \& Technology, University of Maryland, College Park, Maryland 20742, USA}

\author{Samantha Feldman}
\affiliation{Department of Mechanical Engineering, and Institute for Physical Science \& Technology, University of Maryland, College Park, Maryland 20742, USA}
\affiliation{Department of Mechanical Engineering and Materials Science, Washington University in St. Louis, Missouri 63130, USA}

\author{You-Chia Chang}
\affiliation{Department of Photonics, College of Electrical and Computer Engineering, National Yang Ming Chiao Tung University, Hsinchu 30069, Taiwan}

\author{Thomas A. Smith}
\affiliation{Quantum Research and Applications Branch, Naval Air Warfare Center, Patuxent River, Maryland 20670, USA}

\author{Avik Dutt}
\email{avikdutt@umd.edu}
\affiliation{Department of Mechanical Engineering, and Institute for Physical Science \& Technology, University of Maryland, College Park, Maryland 20742, USA}
\affiliation{National Quantum Laboratory (QLab) at Maryland, College Park, Maryland 20740, USA}

\begin{abstract}
Squeezed light, with its quantum noise reduction capabilities, has emerged as a powerful resource in quantum information processing and precision metrology. To reach noise reduction levels such that a quantum advantage is achieved, off-chip squeezers are typically used. The development of on-chip squeezed light sources, particularly in nanophotonic platforms, has been challenging. We report 3.7 ± 0.2 dB of directly detected nanophotonic quantum squeezing using foundry-fabricated silicon nitride (Si$_3$N$_4$) microrings with an inferred squeezing level of 10.7 dB on-chip. The squeezing level is robust across multiple devices and pump detunings, and is consistent with the overcoupling degree without noticeable degradation from excess classical noise. We also offer insights to mitigate thermally-induced excess noise, that typically degrades squeezing, by using small-radius rings with a larger free spectral range (450 GHz) and consequently lower parametric oscillation thresholds. Our results demonstrate that Si$_3$N$_4$ is a viable platform for strong quantum noise reduction in a CMOS-compatible, scalable architecture.
\end{abstract}

\maketitle


\section{Introduction}

Squeezed light exhibits noise in one quadrature reduced below the shot noise, while the conjugate quadrature experiences increased noise as dictated by the Heisenberg uncertainty relation. This quantum noise reduction, quantified by squeezing level (SL), finds myriad applications in continuous-variable (CV) quantum information processing \cite{Braunstein:05} and enhanced precision in metrology and sensing \cite{Lawrie:19}. Quantum sensors leveraging squeezed light have demonstrated enhanced sensitivity in gravitational wave detection \cite{Aasi:13}, spectroscopy \cite{Polzik:1992, deAndrade:20}, quantum imaging\cite{Boyer:08, Andersen:02}, and absolute detector calibration \cite{Marino:11, Vahlbruch:16}. To effectively harness quantum noise reduction for aforementioned applications, it is essential to achieve a large and repeatable degree of squeezing.

In off-chip platforms, significant levels of squeezing have been consistently achieved, both for quadrature-squeezed vacuum and in bright squeezed beams. Notable examples include high levels of squeezing in LiNbO$_3$ crystals \cite{Vahlbruch:08}, rubidium atomic vapor \cite{McCormick:08}, and periodically poled KTP crystals \cite{Mehmet:11, Eberle:08}, leading to the current record of 15 dB \cite{Vahlbruch:16}. The development of vacuum squeezed sources has been used for quantum sensing applications  \cite{Aasi:13, Polzik:1992, Boyer:08, Vahlbruch:16}, as well as for quantum information processing applications such as: the generation of time-multiplexed cluster states \cite{Yokoyama:13, Asavanant:19, Larsen:19}, spectrally multiplexed quantum frequency combs \cite{chen:14, Roslund:14, Kouadou:23}, and for achieving quantum computational advantage \cite{Zhong:20, Madsen:22}. 
Meanwhile, bright squeezing, particularly for the twin-beam mode of operation, has gained significant traction due to setup simplicity that obviates homodyne detection. For sensing, it is also natural to use one of the twin beams to probe samples and the other one as a reference, as has been shown in micro-mechanical and plasmonic sensing \cite{Pooser:15, Dowran:18, Pooser:16}, spectroscopy \cite{deAndrade:20,Li:22}, microscopy \cite{Samantaray:17} and biosensing beyond the photodamage limit \cite{Li:21, Li:24, Casacio:21}. However, the applications of squeezed light have been primarily limited to off-chip platforms.

Compared to bulk optics, in nanophotonics the detected levels of squeezing have been much lower despite a decade-long effort. Nevertheless, nanophotonic squeezers hold promise for practical quantum utility due to smaller device footprints on mass-manufacturable chips. Building upon the advances in bulk optic squeezers, optical parametric oscillators (OPO) and amplifiers (OPA) have been the most commonly used techniques to generate nanophotonic squeezing. In the case of bright (above OPO threshold) squeezed light sources, Fürst \textit{et al.} demonstrate an SL of 2.7 dB using $\chi^{(2)}$ off-chip microresonators \cite{Furst:11}. On-chip Si$_3$N$_4$ microresonators, despite low propagation loss and appreciable $\chi^{(3)}$ nonlinearity, have only shown modest improvements of the SL from 1.7 dB in the initial demonstration \cite{Dutt:15,Dutt:16} to 2.3 dB in recent reports \cite{Kogler:24}. For below-threshold squeezers, silica resonators have demonstrated up to 60 squeezed comb modes with an SL between 1.6 to 3.2 dB \cite{Ze:24, Yang:21} while up to 1.3 dB SL has been presented for integrated vacuum squeezers in Si$_3$N$_4$ and thin-film lithium niobate platforms \cite{Zhao:20, Chen:22, Jahanbozorgi:23,Cernansky:20,Vaidya:20,Stokowski:23,Park:24}.  

One of the major limiting factors in the aforementioned experiments is the loss between the quantum-squeezed light source and the photodetectors. To overcome overall detection loss, thin-film lithium niobate waveguides have recently harnessed an additional phase-sensitive OPA, which converts the squeezed light into a classical, macroscopic state, as discussed in Ref. \cite{Takanashi:20}. The 4.9 dB of squeezing reported using this approach stands as a record in on-chip squeezers, albeit in the pulsed regime \cite{Nehra:22}. Hence, the development of a \textit{continuous-wave} nanophotonic squeezer in a widely accessible, complementary metal-oxide-semiconductor (CMOS)-compatible, and scalable platform such as silicon nitride is desirable \textemdash particularly in the bright twin-beam regime where significant quantum sensing promise has been demonstrated in bulk optics \cite{Pooser:15, Dowran:18, Pooser:16,Li:21, Casacio:21}.

Here we report the largest level of squeezing directly detected from a microresonator, 3.7 ± 0.2 dB in the strongly overcoupled regime (91\% overcoupling), corresponding to 10.7 dB of generated squeezing. This is also a marked improvement over previously detected \textit{continuous-wave} squeezing in nanophotonics \cite{Kogler:24}. With a foundry-fabricated silicon nitride (Si$_3$N$_4$) high-Q microresonator, we demonstrate substantial quantum noise reduction (SL >3 dB) for multiple devices across a range of pump detunings and signal powers, thus ensuring that the observed squeezing is repeatable, long-term stable, and robust against fabrication variations. Additionally, the detected squeezing is commensurate with the degree of overcoupling and detection inefficiency, hence tempering previous concerns and some experiment results about the degradation of two-mode SL from excess classical noise \cite{Dutt:15,Kogler:24,Brusaschi:24}. The achieved squeezing levels open up new possibilities for on-chip quantum-enhanced sensing and information processing applications.

\section{Methods and Experimental Setup}

Consider an optical parametric oscillator based on a second-order or third-order nonlinear process, pumped above threshold. The detected squeezing level (SL) in such a cavity OPO can be modeled by the equation \cite{Fabre:89, Dutt:15, Chembo:16}:
\begin{equation}
\text{SL}\,(\text{dB}) = 10 \log_{10}\left(1 - \frac{\eta_\text{D} \eta_\text{path} \theta}{1+\Omega^2 \tau_c^2}\right)
\label{eq:sl}
\end{equation}
where $\eta_\text{D}$ is the efficiency of the detector, $\eta_\text{path}$ is the propagation efficiency, $\Omega$ is the detection sideband frequency, $\tau_c$ is the cavity photon lifetime, and $\theta$ is the cavity overcoupling coefficient $\theta = 1 - Q_L/Q_i$, with $Q_i$ and $Q_L$ being the intrinsic and loaded quality factors respectively. For bright twin beam squeezing, the squeezing level is nearly independent of parametric gain \cite{Fabre:89}, which contrasts with the gain dependence or equivalently pump-power dependence observed in below-threshold squeezing. For our experiment, we set $\Omega/2\pi = 5$ MHz, which, since $\tau_c<1$ ns, ensured that $\Omega \tau_c \ll 1$. This allows us to ignore the denominator in Eq. \eqref{eq:sl}.

\onecolumngrid

\begin{figure}[htbp]
\centering
\includegraphics[width=0.8\textwidth]{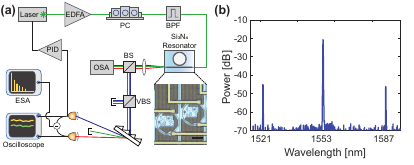}
\caption{(a) Schematic of the measurement setup. OSA: optical spectrum analyzer. PC: polarization controller. BS: beam splitter. VBS: variable polarization beam splitter. EDFA: erbium-doped fiber amplifier. ESA:  electrical spectrum analyzer. We feed the DC output of the signal beam to perform PID control on the laser current (detuning) and stabilize the twin-beam power. A microscope image of the Si$_3$N$_4$ microring resonator with waveguide dimensions of \um{1.3} $\times$ \um{0.78} and a 450-GHz free spectral range. The measured second-order dispersion parameter $D_2$ is 8.6 MHz. Scale bar: \um{50} (b) The optical spectrum of the generated twin beams, which are at 9 FSRs away from the pump. FSR, free spectral range.}
\label{fig:setup}
\end{figure}

\twocolumngrid

\onecolumngrid

\begin{figure}[htbp]
\centering
\includegraphics[width=12cm]{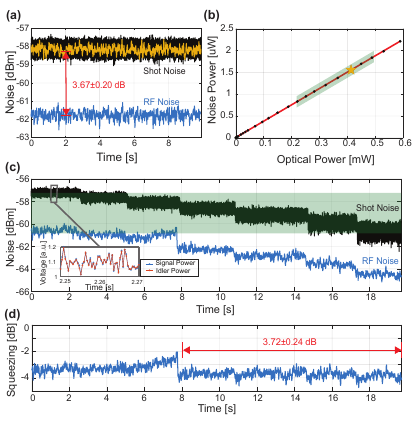}
\caption{(a) Shot noise (yellow) and twin-beam intensity difference noise (blue) taken at 5 MHz measurement frequency, 30 kHz resolution bandwidth (RBW), and 100 Hz video bandwidth (VBW) in zero-span mode on the ESA. Shot noise in black is obtained from the average powers of signal and idler and independent calibration of shot-noise in (b). The dark noise is > 10 dB below the shot noise and has been subtracted from the data. We directly detect a squeezing level of 3.7 dB. (b) Shot noise calibration with dark noise subtracted. The yellow star corresponds to the shot noise in (a). Error bars are smaller than the dots. (c) RF noise of decreasing signal powers (increasing pump detuning), which is adjusted through set points of the feedback loop. The green shaded region in (b) and (c) represents the same signal powers' range. The inset shows the correlated fluctuations of signal and idler power. (d) We detect a constant average squeezing level of 3.7 dB from 8 to 20 s, despite deliberate changes in the signal power during this period, thus showing the robustness of two-mode squeezing above the parametric oscillation threshold. The small deviation during 6-8 s is due to systematic drifts, which are subsequently corrected via electronic feedback. Error bars are determined from the standard deviation of the measured data points.}
\label{fig:RawSq}
\end{figure}

\twocolumngrid

Our quantum light source consists of a Si$_3$N$_4$ point-coupled ring undergoing four-wave mixing (FWM) parametric oscillation. The device was fabricated in a multi-project wafer run by Ligentec  (Fig. \ref{fig:setup}a). A tunable laser (Toptica CTL) is amplified by an erbium-doped fiber amplifier (EDFA) to pump the ring slightly above oscillation threshold to generate quantum correlated twin beams consisting of a lower wavelength signal and a higher wavelength idler (Fig. \ref{fig:setup}b). To mitigate the impact of EDFA noise on the correlation of the twin beams, we employ a 7 nm band-pass filter (BPF). Light is coupled into the chip with a lensed fiber and collected by an antireflection-coated aspheric lens. We use a beam splitter to send 1.5\% of the output to an optical spectrum analyzer (OSA). The remaining light passes through a variable polarization beam splitter (VBS, 94\% measured maximum efficiency) which adjusts the detection path loss $1-\eta_\text{path}$ (Fig. \ref{fig:setup}a) {and ensures that the fundamental transverse-electric mode in the waveguide is excited with the aid of a polarization controller (PC)}. After spatially separating the twin beams and the pump (blocked) with a transmission grating, we focus the twin beams onto a balanced detector ($\eta_\text{D} = 79\%$). The measured $\eta_\text{path}$ is 77\%, resulting from the optics components loss in the detection path. The differential RF output of the detector is sent to an electrical spectrum analyzer (ESA). The DC outputs of the detector are sent to an oscilloscope to monitor the twin beam powers, from which the shot noise can be obtained using an independent calibration (Fig. \ref{fig:RawSq}b). We use a PID feedback loop actuating on the laser current to lock one of the twin beam powers at a constant set point, which effectively stabilizes the pump detuning. 

\section{Experimental Results}

\begin{figure}[htbp]
\centering
\includegraphics[width=12cm]{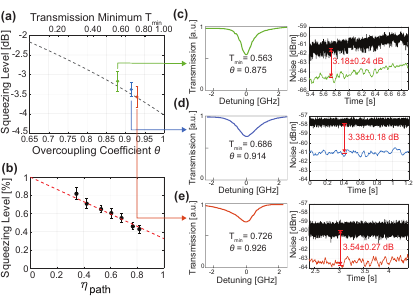}
\caption{(a) Measured squeezing level versus overcoupling coefficient $\theta$ of 0.87 (green, without locking), 0.91 (blue, locked), and 0.93 (orange, locked). The blue and orange points are taken using the same device with a ring-bus gap of \um{0.3}, but at two pump resonances (1553.5 nm and 1560.6 nm). The green point is taken at a resonance of 1553.5 nm with a ring bus gap of \um{0.35}. The dashed line shows the fitted trend in SL vs. $\theta$ (Eq. 1) for $\eta_\text{path}=77\%$ from a known value of $\eta_\text{D}=79\%$ and $\theta$ deduced from Lorentzian fits. The spectral dependence of $\eta_\text{D}$ and the grating efficiency are not accounted for. (c-e) Linear transmission measurement from which we extract $\theta$ and the corresponding intensity difference noise measurement. (b) Verification of the linear degradation of squeezing with deliberately introduced attenuation of $\eta_\text{path}$, approaching the shot noise as $\eta_\text{path}\to 0$. All data points in (a) and (b) are taken with the VBS in the detection path, except for the rightmost point in (b) for which $\eta_\text{path} = 81\%$ without VBS, which also corresponds to the measurement shown in Fig. \ref{fig:RawSq}. }
\label{fig:SqOverC}
\end{figure}

{We observe a stable squeezed state with an average squeezing level of 3.7 dB, as shown in Fig. \ref{fig:RawSq}a. We obtain the shot noise (black line) by performing an independent calibration on the balanced detector. For this calibration, two beams with identical optical power from the laser, ranging from \uw{2.3} up to \uw{590}, are sent to the two ports of the balanced detector. The resulting differential RF outputs are recorded by an ESA with a resolution bandwidth (RBW) of 30 kHz and a video bandwidth (VBW) of 100 Hz in zero-span mode at $\Omega/2\pi = 5$ MHz. Figure \ref{fig:RawSq}b shows this shot noise, with dark noise subtracted, as a function of input optical power, which is linear over the 24-dBm optical power span.
    When twin beams are generated, we performed intensity difference noise measurements on the ESA with the same setting, continuously recording 50-ms scans containing 751 data points, while twin beam powers are simultaneously acquired by an oscilloscope. The ESA trace in yellow shows the shot noise level corresponding to the scenario where both ports of the balanced detector received the same input power of \uw{410}, corresponding to the average twin beam powers when the intensity difference noise in (Fig. \ref{fig:RawSq}a) were acquired. To obtain the best detection efficiency, the VBS is not in the detection path for measurements in Fig. \ref{fig:RawSq}, which results in a $\eta_{path}=81\%$. As previously mentioned, the PID feedback loop enables us to maintain steady twin beam powers and a constant squeezing level (Fig. \ref{fig:RawSq}a) in the entire 10-second trace, as opposed to drifts in the signal power for a free-running setup, an example of which is shown later in Fig. \ref{fig:SqOverC}c.}

Next, we demonstrate the robustness of above-threshold twin-beam squeezing despite deliberate changes in the signal power and the pump detuning (Fig. \ref{fig:RawSq}c,d). We lock one of the twin beams (the signal power) at monotonically decreasing discrete values, which effectively moves the pump laser’s detuning further away from the microring resonance at 1553.5 nm, as shown in Fig. \ref{fig:RawSq}c. The blue squeezing traces in Fig. \ref{fig:RawSq}c and \ref{fig:RawSq}d show a decrease in the squeezing level between 6 to 8 s, followed by a recovery from the perturbation, maintaining it at a consistent level of $3.7 \pm 0.2$ dB for the rest of the measurement. The same SL between 8 to 20 seconds indicates that small-change in pump detuning (<100 MHz) near resonance does not affect the squeezing level. In practice, we observe the squeezing remains stable on timescales of a few minutes without noticeable influence from laser noise and mechanical fluctuations.

\onecolumngrid

\twocolumngrid

The measured squeezing is consistent with the overcoupling coefficient $\theta$ and the optical path loss $\eta_{\rm path}$ in accordance with Eq. \eqref{eq:sl}, as shown in Fig. \ref{fig:SqOverC}a.
The overcoupling coefficient $\theta = 1-Q_L/Q_i$ is calculated from a Lorentzian fit of the linear transmission through the microring which provides the loaded quality factor $Q_L$ and the intrinsic quality factor $Q_i$. The $Q$ measurements were taken at low optical powers below \uw{100} to avoid nonlinear broadening of the resonance. 
After calculating $\theta$ for a few different resonances and ring-bus coupling gaps, we fit the measured squeezing level using $\eta_{\rm path}$ as the only free parameter, which results in the dashed gray line in Fig. \ref{fig:SqOverC}a. The detection efficiency $\eta_D$ was known \textit{a priori} to be 79\%. 
The measured squeezing level agrees well with the dashed gray line for fitted $\eta_{\rm path}=77\%$, which 
matches the independently measured $\eta_\text{path}=77\%$, much in the spirit of absolute detector calibration that has been predicted \cite{Fabre:89} and demonstrated \cite{Marino:11} using squeezed light. This indicates that there is no measurable impact of excess intensity noise that diminishes the twin-beam squeezing level. As a further check, we verified that the twin-beam intensity-difference noise linearly approaches the shot noise level as we intentionally added loss to the system, thereby reducing $\eta_\text{path}$ and mixing the squeezed state with vacuum (Fig. \ref{fig:SqOverC}b). All these independent measurements and calibrations provide strong evidence in favor of an inferred on-chip squeezing level of 11.5 dB for the best result (orange point in Fig. \ref{fig:SqOverC}a) with $\theta$ = 0.93. 

\onecolumngrid

\begin{figure}[htbp]
\centering
\includegraphics[width=12cm]{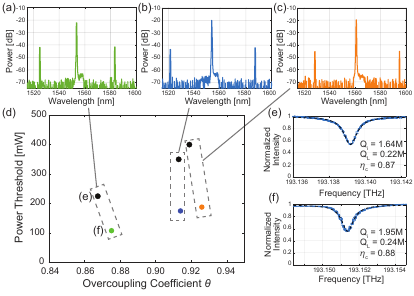}
\caption{(d) Variation of the optical parametric oscillation power threshold with the overcoupling coefficienct $\theta$. Each dashed box represents the same gap and waveguide dimensions, comparing larger (450 GHz, colored) and smaller (210 GHz, black) FSRs. (a-c) Optical spectra of the twin beams generated at 8 FSRs (a) and 9 FSRs (b, c) away from the pump, corresponding to the three squeezing measurements in Fig. \ref{fig:SqOverC}a from the 450 GHz FSR rings. The signal and idler wavelengths generated for the smaller FSR ring (210 GHz, black dots) are within 0.4 nm from the signal and idler wavelengths for the corresponding 450 GHz FSR rings, and are hence not plotted here. (e-f) Linear transmission measurements and the fitted value of $Q_i$ and $Q_c$ for different size rings.}
\label{fig:threshold}
\end{figure}

\twocolumngrid

During our measurements, we observed that a large FSR ring of 450 GHz provided advantages for bright twin-beam squeezing due to the lower OPO power threshold. The threshold power for FWM parametric oscillation can be approximated as \cite{Ji:17,Dutt:24, hausmann_diamond_2014}:
\begin{equation}
P_{\rm th} \approx 1.54 \frac{\pi}{4 \theta} \frac{c n_0}{n_2 \lambda_{\rm p}} \frac{A_{\rm eff}}{{\rm FSR}\cdot Q_L^2}
\end{equation}  
where $n_0 (n_2)$ is the linear (nonlinear) refractive index, $\lambda_p$ is the pump wavelength, and $A_{\rm eff}$ is the effective mode area.
We see that $P_{\rm th}$ is inversely proportional to the resonator's FSR for a given waveguide dimension. In Fig. \ref{fig:threshold}, we plot the measured oscillation threshold for each of the  squeezing measurements reported earlier in Fig. \ref{fig:SqOverC}, using colored dots. These threshold powers are compared with the threshold power for corresponding microrings with a smaller FSR (210 GHz) but identical waveguide dimensions and ring-bus coupling gaps, represented by black dots in Fig. \ref{fig:threshold}d. For the same ring-bus coupling gap, the measured $\theta$ is similar, while we observed a 2-2.5 times lower $P_{\rm th}$ for microrings with a 450 GHz FSR than those with a 210 GHz FSR. This significant reduction in threshold power with larger FSR rings reduces the power on the fiber-to-chip coupling setup as well as the intracavity power, which alleviates thermal-induced drifts, enhancing the stability of the optical parametric oscillators. In addition, a larger FSR provides a larger window over which the mode number separation between the generated twin beams remains the same \cite{Herr:12}.

Several Si$_3$N$_4$ squeezers in existing literature have observed lower levels of quantum noise reduction than that predicted by Eq. \eqref{eq:sl} and the mismatch was attributed to thermally induced noise or laser phase noise \cite{Cernansky:20, Dutt:15, Kogler:24, Brusaschi:24}. Our results, particularly those in Fig. \ref{fig:SqOverC}, provide a contrasting picture, at least for twin-beam intensity difference squeezing generated in silicon nitride. 
Specifically, Kogler \textit{et al.} and Brusaschi \textit{et al.} inferred 5 dB and 6.2 dB on-chip squeezing respectively, compared to 9 dB and 11 dB expected from their devices' overcoupling. Beyond twin-beam squeezing in FWM OPOs, thermorefractive noise has also been suspected to play a role in self-phase-modulation-based squeezers in Si$_3$N$_4$, limiting the the observed squeezing to 0.45 dB at high RF detection frequency \cite{Cernansky:20} and no squeezing below 500 MHz. All three attribute this to unanticipated parasitic thermally-induced noise processes that contaminate the squeezed state. Similarly, we did not observe squeezing that is commensurate with $\theta$ from devices with a threshold power exceeding 250 mW (210 GHz rings in Fig. \ref{fig:threshold}) at the RF detection frequency used in our experiments. Even for the 450-GHz FSR ring, the stable squeezing measurement in Fig. \ref{fig:RawSq} was not seen for the most overcoupled device ($\theta = 0.93$), likely attributable to the increased excess noise associated with higher pump power. Nevertheless, Fig.\ref{fig:SqOverC}a does show that, in smaller rings, excess classical noise is mitigated by the $\sim$ 35 dB common-mode rejection ratio of the balanced detection system at lower pump power thresholds. These findings suggest that for generating squeezed state in Kerr microresonator OPOs, it is favorable to use smaller rings as long as the bending loss originating from a tighter bend radius in such small rings does not dominate the scattering loss. We verified that our devices indeed operate in this regime as we measured no systematic decrease in $Q_i$ for the smaller radius, 450-GHz FSR rings compared to larger radius, 210-GHz FSR rings.

\section{Discussion}
We report 3.7 dB of quantum noise reduction detected from a Si$_3$N$_4$ microresonator, which represents a marked improvement over previous microresonator-based squeezed light sources. Although the measured SL is lower than off-chip systems, importantly, our Si$_3$N$_4$ microresonators can generate intensity-difference squeezing that matches the expected SL based on the overcoupling coefficient. In off-chip optical parametric oscillators where 15-dB of strong squeezing was observed \cite{Vahlbruch:16}, the total detection efficiency has been optimized to 98.5\%, while our current setup achieves an overall maximum detection efficiency ($\eta_D \eta_{path}$) of 64\%. Given the minimal excess noise and large inferred value of squeezing on-chip for Si$_3$N$_4$ squeezers, this gap can be bridged by further reducing chip out-coupling loss and employing a higher quantum efficiency balanced detector. On the other hand, several sensing applications \cite{Li:21} can take advantage of the on-chip generated high squeezing level $>$10 dB if the samples to be sensed are integrated on the same chip.  

These results provide a strong foundation for further development of integrated quantum photonic devices, including squeezers, filters, and balanced photodetection modules with CMOS electronics to significantly reduce total device footprint without compromising loss from coupling light out of the chip. In pursuit of this objective, high extinction passive and actively tunable filters have been demonstrated in silicon photonics \cite{Huffman:17, Nie:19, Afifi:21, Singh:22}. Regarding integrated balanced-detection, significant improvements have been recently reported for bandwidth and common-mode rejection ratio (CMRR) through electronic-photonic integration \cite{Gao:23,Costanzo:21,Bruynsteen:21,Raffaelli:18,Tasker:24}; however, achieving high external quantum efficiency to detect squeezing remains a challenge \cite{Tasker:21,Gurses:24}. Beyond silicon nitride, other $\chi^{(2)}$ and $\chi^{(3)}$ nonlinear optical materials such as AlGaAs, InGaP, diamond and silicon carbide have made progress in low-loss on-chip integration, and they promise to be fruitful for generating squeezed light \cite{lustig_emerging_2024, pu_ultra-efficient_2018, xie_ultrahigh-q_2020, hausmann_diamond_2014}. 
In light of these complementary advances, our work represents a significant step towards realizing high-performance, compact quantum photonic devices, paving the way for advanced quantum sensing, communication, and computation applications \cite{Guidry:23,Dutt:24,pontula_2024_multimode, moody_2022_2022}.

\section*{Funding}
National Science Foundation (2326792), NAWCAD (N004212310002).



\section*{Acknowledgment}
We acknowledge Y. Chembo for providing essential equipment, and P. D. Lett and Z. Zhou for insightful discussions.

\bibliography{sample}

\begin{thebibliography}{68}%
\makeatletter
\providecommand \@ifxundefined [1]{%
 \@ifx{#1\undefined}
}%
\providecommand \@ifnum [1]{%
 \ifnum #1\expandafter \@firstoftwo
 \else \expandafter \@secondoftwo
 \fi
}%
\providecommand \@ifx [1]{%
 \ifx #1\expandafter \@firstoftwo
 \else \expandafter \@secondoftwo
 \fi
}%
\providecommand \natexlab [1]{#1}%
\providecommand \enquote  [1]{``#1''}%
\providecommand \bibnamefont  [1]{#1}%
\providecommand \bibfnamefont [1]{#1}%
\providecommand \citenamefont [1]{#1}%
\providecommand \href@noop [0]{\@secondoftwo}%
\providecommand \href [0]{\begingroup \@sanitize@url \@href}%
\providecommand \@href[1]{\@@startlink{#1}\@@href}%
\providecommand \@@href[1]{\endgroup#1\@@endlink}%
\providecommand \@sanitize@url [0]{\catcode `\\12\catcode `\$12\catcode `\&12\catcode `\#12\catcode `\^12\catcode `\_12\catcode `\%12\relax}%
\providecommand \@@startlink[1]{}%
\providecommand \@@endlink[0]{}%
\providecommand \url  [0]{\begingroup\@sanitize@url \@url }%
\providecommand \@url [1]{\endgroup\@href {#1}{\urlprefix }}%
\providecommand \urlprefix  [0]{URL }%
\providecommand \Eprint [0]{\href }%
\providecommand \doibase [0]{https://doi.org/}%
\providecommand \selectlanguage [0]{\@gobble}%
\providecommand \bibinfo  [0]{\@secondoftwo}%
\providecommand \bibfield  [0]{\@secondoftwo}%
\providecommand \translation [1]{[#1]}%
\providecommand \BibitemOpen [0]{}%
\providecommand \bibitemStop [0]{}%
\providecommand \bibitemNoStop [0]{.\EOS\space}%
\providecommand \EOS [0]{\spacefactor3000\relax}%
\providecommand \BibitemShut  [1]{\csname bibitem#1\endcsname}%
\let\auto@bib@innerbib\@empty
\bibitem [{\citenamefont {Braunstein}\ and\ \citenamefont {van Loock}(2005)}]{Braunstein:05}%
  \BibitemOpen
  \bibfield  {author} {\bibinfo {author} {\bibfnamefont {S.~L.}\ \bibnamefont {Braunstein}}\ and\ \bibinfo {author} {\bibfnamefont {P.}~\bibnamefont {van Loock}},\ }\bibfield  {title} {\bibinfo {title} {Quantum information with continuous variables},\ }\href@noop {} {\bibfield  {journal} {\bibinfo  {journal} {Rev. Mod. Phys.}\ }\textbf {\bibinfo {volume} {77}} (\bibinfo {year} {2005})}\BibitemShut {NoStop}%
\bibitem [{\citenamefont {Lawrie}\ \emph {et~al.}(2019)\citenamefont {Lawrie}, \citenamefont {Lett}, \citenamefont {Marino},\ and\ \citenamefont {Pooser}}]{Lawrie:19}%
  \BibitemOpen
  \bibfield  {author} {\bibinfo {author} {\bibfnamefont {B.~J.}\ \bibnamefont {Lawrie}}, \bibinfo {author} {\bibfnamefont {P.~D.}\ \bibnamefont {Lett}}, \bibinfo {author} {\bibfnamefont {A.~M.}\ \bibnamefont {Marino}},\ and\ \bibinfo {author} {\bibfnamefont {R.~C.}\ \bibnamefont {Pooser}},\ }\bibfield  {title} {\bibinfo {title} {Quantum sensing with squeezed light},\ }\href@noop {} {\bibfield  {journal} {\bibinfo  {journal} {ACS Photonics}\ }\textbf {\bibinfo {volume} {6}},\ \bibinfo {pages} {1307} (\bibinfo {year} {2019})}\BibitemShut {NoStop}%
\bibitem [{\citenamefont {Aasi}\ \emph {et~al.}(2013)\citenamefont {Aasi}, \citenamefont {Abadie}, \citenamefont {Abbott} \emph {et~al.}}]{Aasi:13}%
  \BibitemOpen
  \bibfield  {author} {\bibinfo {author} {\bibfnamefont {J.}~\bibnamefont {Aasi}}, \bibinfo {author} {\bibfnamefont {J.}~\bibnamefont {Abadie}}, \bibinfo {author} {\bibfnamefont {B.}~\bibnamefont {Abbott}}, \emph {et~al.},\ }\bibfield  {title} {\bibinfo {title} {Enhanced sensitivity of the ligo gravitational wave detector by using squeezed states of light},\ }\href {https://doi.org/10.1038/nphoton.2013.177} {\bibfield  {journal} {\bibinfo  {journal} {Nature Photonics}\ }\textbf {\bibinfo {volume} {7}},\ \bibinfo {pages} {613} (\bibinfo {year} {2013})}\BibitemShut {NoStop}%
\bibitem [{\citenamefont {Polzik}\ \emph {et~al.}(1992)\citenamefont {Polzik}, \citenamefont {Carri},\ and\ \citenamefont {Kimble}}]{Polzik:1992}%
  \BibitemOpen
  \bibfield  {author} {\bibinfo {author} {\bibfnamefont {E.~S.}\ \bibnamefont {Polzik}}, \bibinfo {author} {\bibfnamefont {J.}~\bibnamefont {Carri}},\ and\ \bibinfo {author} {\bibfnamefont {H.~J.}\ \bibnamefont {Kimble}},\ }\bibfield  {title} {\bibinfo {title} {Spectroscopy with squeezed light},\ }\href {https://doi.org/10.1103/PhysRevLett.68.3020} {\bibfield  {journal} {\bibinfo  {journal} {Phys. Rev. Lett.}\ }\textbf {\bibinfo {volume} {68}},\ \bibinfo {pages} {3020} (\bibinfo {year} {1992})}\BibitemShut {NoStop}%
\bibitem [{\citenamefont {de~Andrade}\ \emph {et~al.}(2020)\citenamefont {de~Andrade}, \citenamefont {Kerdoncuff}, \citenamefont {Berg-S{\o}rensen}, \citenamefont {Gehring}, \citenamefont {Lassen},\ and\ \citenamefont {Andersen}}]{deAndrade:20}%
  \BibitemOpen
  \bibfield  {author} {\bibinfo {author} {\bibfnamefont {R.~B.}\ \bibnamefont {de~Andrade}}, \bibinfo {author} {\bibfnamefont {H.}~\bibnamefont {Kerdoncuff}}, \bibinfo {author} {\bibfnamefont {K.}~\bibnamefont {Berg-S{\o}rensen}}, \bibinfo {author} {\bibfnamefont {T.}~\bibnamefont {Gehring}}, \bibinfo {author} {\bibfnamefont {M.}~\bibnamefont {Lassen}},\ and\ \bibinfo {author} {\bibfnamefont {U.~L.}\ \bibnamefont {Andersen}},\ }\bibfield  {title} {\bibinfo {title} {Quantum-enhanced continuous-wave stimulated raman scattering spectroscopy},\ }\href {https://doi.org/10.1364/OPTICA.386584} {\bibfield  {journal} {\bibinfo  {journal} {Optica}\ }\textbf {\bibinfo {volume} {7}},\ \bibinfo {pages} {470} (\bibinfo {year} {2020})}\BibitemShut {NoStop}%
\bibitem [{\citenamefont {Boyer}\ \emph {et~al.}(2008)\citenamefont {Boyer} \emph {et~al.}}]{Boyer:08}%
  \BibitemOpen
  \bibfield  {author} {\bibinfo {author} {\bibfnamefont {V.}~\bibnamefont {Boyer}} \emph {et~al.},\ }\bibfield  {title} {\bibinfo {title} {Entangled images from four-wave mixing},\ }\href {https://doi.org/10.1126/science.1158275} {\bibfield  {journal} {\bibinfo  {journal} {Science}\ }\textbf {\bibinfo {volume} {321}},\ \bibinfo {pages} {544} (\bibinfo {year} {2008})}\BibitemShut {NoStop}%
\bibitem [{\citenamefont {Treps}\ \emph {et~al.}(2002)\citenamefont {Treps}, \citenamefont {Andersen}, \citenamefont {Buchler}, \citenamefont {Lam}, \citenamefont {Ma\^{\i}tre}, \citenamefont {Bachor},\ and\ \citenamefont {Fabre}}]{Andersen:02}%
  \BibitemOpen
  \bibfield  {author} {\bibinfo {author} {\bibfnamefont {N.}~\bibnamefont {Treps}}, \bibinfo {author} {\bibfnamefont {U.}~\bibnamefont {Andersen}}, \bibinfo {author} {\bibfnamefont {B.}~\bibnamefont {Buchler}}, \bibinfo {author} {\bibfnamefont {P.~K.}\ \bibnamefont {Lam}}, \bibinfo {author} {\bibfnamefont {A.}~\bibnamefont {Ma\^{\i}tre}}, \bibinfo {author} {\bibfnamefont {H.-A.}\ \bibnamefont {Bachor}},\ and\ \bibinfo {author} {\bibfnamefont {C.}~\bibnamefont {Fabre}},\ }\bibfield  {title} {\bibinfo {title} {Surpassing the standard quantum limit for optical imaging using nonclassical multimode light},\ }\href {https://doi.org/10.1103/PhysRevLett.88.203601} {\bibfield  {journal} {\bibinfo  {journal} {Phys. Rev. Lett.}\ }\textbf {\bibinfo {volume} {88}},\ \bibinfo {pages} {203601} (\bibinfo {year} {2002})}\BibitemShut {NoStop}%
\bibitem [{\citenamefont {Marino}\ and\ \citenamefont {Lett}(2011)}]{Marino:11}%
  \BibitemOpen
  \bibfield  {author} {\bibinfo {author} {\bibfnamefont {A.~M.}\ \bibnamefont {Marino}}\ and\ \bibinfo {author} {\bibfnamefont {P.~D.}\ \bibnamefont {Lett}},\ }\bibfield  {title} {\bibinfo {title} {Absolute calibration of photodiodes with bright twin beams},\ }\href@noop {} {\bibfield  {journal} {\bibinfo  {journal} {Journal of Modern Optics}\ }\textbf {\bibinfo {volume} {58}},\ \bibinfo {pages} {328} (\bibinfo {year} {2011})}\BibitemShut {NoStop}%
\bibitem [{\citenamefont {Vahlbruch}\ \emph {et~al.}(2016)\citenamefont {Vahlbruch}, \citenamefont {Mehmet}, \citenamefont {Danzmann},\ and\ \citenamefont {Schnabel}}]{Vahlbruch:16}%
  \BibitemOpen
  \bibfield  {author} {\bibinfo {author} {\bibfnamefont {H.}~\bibnamefont {Vahlbruch}}, \bibinfo {author} {\bibfnamefont {M.}~\bibnamefont {Mehmet}}, \bibinfo {author} {\bibfnamefont {K.}~\bibnamefont {Danzmann}},\ and\ \bibinfo {author} {\bibfnamefont {R.}~\bibnamefont {Schnabel}},\ }\bibfield  {title} {\bibinfo {title} {Detection of 15 db squeezed states of light and their application for the absolute calibration of photoelectric quantum efficiency},\ }\href@noop {} {\bibfield  {journal} {\bibinfo  {journal} {Phys. Rev. Lett.}\ }\textbf {\bibinfo {volume} {117}},\ \bibinfo {pages} {110801} (\bibinfo {year} {2016})}\BibitemShut {NoStop}%
\bibitem [{\citenamefont {Vahlbruch}\ \emph {et~al.}(2008)\citenamefont {Vahlbruch}, \citenamefont {Mehmet}, \citenamefont {Chelkowski}, \citenamefont {Hage}, \citenamefont {Franzen}, \citenamefont {Lastzka}, \citenamefont {Go\ss{}ler}, \citenamefont {Danzmann},\ and\ \citenamefont {Schnabel}}]{Vahlbruch:08}%
  \BibitemOpen
  \bibfield  {author} {\bibinfo {author} {\bibfnamefont {H.}~\bibnamefont {Vahlbruch}}, \bibinfo {author} {\bibfnamefont {M.}~\bibnamefont {Mehmet}}, \bibinfo {author} {\bibfnamefont {S.}~\bibnamefont {Chelkowski}}, \bibinfo {author} {\bibfnamefont {B.}~\bibnamefont {Hage}}, \bibinfo {author} {\bibfnamefont {A.}~\bibnamefont {Franzen}}, \bibinfo {author} {\bibfnamefont {N.}~\bibnamefont {Lastzka}}, \bibinfo {author} {\bibfnamefont {S.}~\bibnamefont {Go\ss{}ler}}, \bibinfo {author} {\bibfnamefont {K.}~\bibnamefont {Danzmann}},\ and\ \bibinfo {author} {\bibfnamefont {R.}~\bibnamefont {Schnabel}},\ }\bibfield  {title} {\bibinfo {title} {Observation of squeezed light with 10-db quantum-noise reduction},\ }\href {https://doi.org/10.1103/PhysRevLett.100.033602} {\bibfield  {journal} {\bibinfo  {journal} {Phys. Rev. Lett.}\ }\textbf {\bibinfo {volume} {100}},\ \bibinfo {pages} {033602} (\bibinfo {year} {2008})}\BibitemShut {NoStop}%
\bibitem [{\citenamefont {McCormick}\ \emph {et~al.}(2008)\citenamefont {McCormick}, \citenamefont {Marino}, \citenamefont {Boyer},\ and\ \citenamefont {Lett}}]{McCormick:08}%
  \BibitemOpen
  \bibfield  {author} {\bibinfo {author} {\bibfnamefont {C.~F.}\ \bibnamefont {McCormick}}, \bibinfo {author} {\bibfnamefont {A.~M.}\ \bibnamefont {Marino}}, \bibinfo {author} {\bibfnamefont {V.}~\bibnamefont {Boyer}},\ and\ \bibinfo {author} {\bibfnamefont {P.~D.}\ \bibnamefont {Lett}},\ }\bibfield  {title} {\bibinfo {title} {Strong low-frequency quantum correlations from a four-wave-mixing amplifier},\ }\href {https://doi.org/10.1103/PhysRevA.78.043816} {\bibfield  {journal} {\bibinfo  {journal} {Phys. Rev. A}\ }\textbf {\bibinfo {volume} {78}},\ \bibinfo {pages} {043816} (\bibinfo {year} {2008})}\BibitemShut {NoStop}%
\bibitem [{\citenamefont {Mehmet}\ \emph {et~al.}(2011)\citenamefont {Mehmet}, \citenamefont {Ast}, \citenamefont {Eberle}, \citenamefont {Steinlechner}, \citenamefont {Vahlbruch},\ and\ \citenamefont {Schnabel}}]{Mehmet:11}%
  \BibitemOpen
  \bibfield  {author} {\bibinfo {author} {\bibfnamefont {M.}~\bibnamefont {Mehmet}}, \bibinfo {author} {\bibfnamefont {S.}~\bibnamefont {Ast}}, \bibinfo {author} {\bibfnamefont {T.}~\bibnamefont {Eberle}}, \bibinfo {author} {\bibfnamefont {S.}~\bibnamefont {Steinlechner}}, \bibinfo {author} {\bibfnamefont {H.}~\bibnamefont {Vahlbruch}},\ and\ \bibinfo {author} {\bibfnamefont {R.}~\bibnamefont {Schnabel}},\ }\bibfield  {title} {\bibinfo {title} {Squeezed light at 1550 nm with a quantum noise reduction of 12.3 db},\ }\href {https://doi.org/10.1364/OE.19.025763} {\bibfield  {journal} {\bibinfo  {journal} {Opt. Express}\ }\textbf {\bibinfo {volume} {19}},\ \bibinfo {pages} {25763} (\bibinfo {year} {2011})}\BibitemShut {NoStop}%
\bibitem [{\citenamefont {Eberle}\ \emph {et~al.}(2010)\citenamefont {Eberle}, \citenamefont {Steinlechner}, \citenamefont {Bauchrowitz}, \citenamefont {H\"andchen}, \citenamefont {Vahlbruch}, \citenamefont {Mehmet}, \citenamefont {M\"uller-Ebhardt},\ and\ \citenamefont {Schnabel}}]{Eberle:08}%
  \BibitemOpen
  \bibfield  {author} {\bibinfo {author} {\bibfnamefont {T.}~\bibnamefont {Eberle}}, \bibinfo {author} {\bibfnamefont {S.}~\bibnamefont {Steinlechner}}, \bibinfo {author} {\bibfnamefont {J.}~\bibnamefont {Bauchrowitz}}, \bibinfo {author} {\bibfnamefont {V.}~\bibnamefont {H\"andchen}}, \bibinfo {author} {\bibfnamefont {H.}~\bibnamefont {Vahlbruch}}, \bibinfo {author} {\bibfnamefont {M.}~\bibnamefont {Mehmet}}, \bibinfo {author} {\bibfnamefont {H.}~\bibnamefont {M\"uller-Ebhardt}},\ and\ \bibinfo {author} {\bibfnamefont {R.}~\bibnamefont {Schnabel}},\ }\bibfield  {title} {\bibinfo {title} {Quantum enhancement of the zero-area sagnac interferometer topology for gravitational wave detection},\ }\href {https://doi.org/10.1103/PhysRevLett.104.251102} {\bibfield  {journal} {\bibinfo  {journal} {Phys. Rev. Lett.}\ }\textbf {\bibinfo {volume} {104}},\ \bibinfo {pages} {251102} (\bibinfo {year} {2010})}\BibitemShut {NoStop}%
\bibitem [{\citenamefont {Yokoyama}\ \emph {et~al.}(2013)\citenamefont {Yokoyama}, \citenamefont {Ukai}, \citenamefont {Armstrong} \emph {et~al.}}]{Yokoyama:13}%
  \BibitemOpen
  \bibfield  {author} {\bibinfo {author} {\bibfnamefont {S.}~\bibnamefont {Yokoyama}}, \bibinfo {author} {\bibfnamefont {R.}~\bibnamefont {Ukai}}, \bibinfo {author} {\bibfnamefont {S.}~\bibnamefont {Armstrong}}, \emph {et~al.},\ }\bibfield  {title} {\bibinfo {title} {Ultra-large-scale continuous-variable cluster states multiplexed in the time domain},\ }\href {https://doi.org/10.1038/nphoton.2013.287} {\bibfield  {journal} {\bibinfo  {journal} {Nature Photonics}\ }\textbf {\bibinfo {volume} {7}},\ \bibinfo {pages} {982} (\bibinfo {year} {2013})}\BibitemShut {NoStop}%
\bibitem [{\citenamefont {Asavanant}\ \emph {et~al.}(2019)\citenamefont {Asavanant}, \citenamefont {Shiozawa}, \citenamefont {Yokoyama}, \citenamefont {Charoensombutamon}, \citenamefont {Emura}, \citenamefont {Alexander}, \citenamefont {Takeda}, \citenamefont {ichi Yoshikawa}, \citenamefont {Menicucci}, \citenamefont {Yonezawa},\ and\ \citenamefont {Furusawa}}]{Asavanant:19}%
  \BibitemOpen
  \bibfield  {author} {\bibinfo {author} {\bibfnamefont {W.}~\bibnamefont {Asavanant}}, \bibinfo {author} {\bibfnamefont {Y.}~\bibnamefont {Shiozawa}}, \bibinfo {author} {\bibfnamefont {S.}~\bibnamefont {Yokoyama}}, \bibinfo {author} {\bibfnamefont {B.}~\bibnamefont {Charoensombutamon}}, \bibinfo {author} {\bibfnamefont {H.}~\bibnamefont {Emura}}, \bibinfo {author} {\bibfnamefont {R.~N.}\ \bibnamefont {Alexander}}, \bibinfo {author} {\bibfnamefont {S.}~\bibnamefont {Takeda}}, \bibinfo {author} {\bibfnamefont {J.}~\bibnamefont {ichi Yoshikawa}}, \bibinfo {author} {\bibfnamefont {N.~C.}\ \bibnamefont {Menicucci}}, \bibinfo {author} {\bibfnamefont {H.}~\bibnamefont {Yonezawa}},\ and\ \bibinfo {author} {\bibfnamefont {A.}~\bibnamefont {Furusawa}},\ }\bibfield  {title} {\bibinfo {title} {Generation of time-domain-multiplexed two-dimensional cluster state},\ }\href {https://doi.org/10.1126/science.aay2645} {\bibfield  {journal} {\bibinfo  {journal} {Science}\ }\textbf {\bibinfo {volume} {366}},\ \bibinfo {pages}
  {373} (\bibinfo {year} {2019})}\BibitemShut {NoStop}%
\bibitem [{\citenamefont {Larsen}\ \emph {et~al.}(2019)\citenamefont {Larsen}, \citenamefont {Guo}, \citenamefont {Breum}, \citenamefont {Neergaard-Nielsen},\ and\ \citenamefont {Andersen}}]{Larsen:19}%
  \BibitemOpen
  \bibfield  {author} {\bibinfo {author} {\bibfnamefont {M.~V.}\ \bibnamefont {Larsen}}, \bibinfo {author} {\bibfnamefont {X.}~\bibnamefont {Guo}}, \bibinfo {author} {\bibfnamefont {C.~R.}\ \bibnamefont {Breum}}, \bibinfo {author} {\bibfnamefont {J.~S.}\ \bibnamefont {Neergaard-Nielsen}},\ and\ \bibinfo {author} {\bibfnamefont {U.~L.}\ \bibnamefont {Andersen}},\ }\bibfield  {title} {\bibinfo {title} {Deterministic generation of a two-dimensional cluster state},\ }\href {https://doi.org/10.1126/science.aay4354} {\bibfield  {journal} {\bibinfo  {journal} {Science}\ }\textbf {\bibinfo {volume} {366}},\ \bibinfo {pages} {369} (\bibinfo {year} {2019})}\BibitemShut {NoStop}%
\bibitem [{\citenamefont {Chen}\ \emph {et~al.}(2014)\citenamefont {Chen}, \citenamefont {Menicucci},\ and\ \citenamefont {Pfister}}]{chen:14}%
  \BibitemOpen
  \bibfield  {author} {\bibinfo {author} {\bibfnamefont {M.}~\bibnamefont {Chen}}, \bibinfo {author} {\bibfnamefont {N.~C.}\ \bibnamefont {Menicucci}},\ and\ \bibinfo {author} {\bibfnamefont {O.}~\bibnamefont {Pfister}},\ }\bibfield  {title} {\bibinfo {title} {Experimental realization of multipartite entanglement of 60 modes of a quantum optical frequency comb},\ }\href {https://doi.org/10.1103/PhysRevLett.112.120505} {\bibfield  {journal} {\bibinfo  {journal} {Phys. Rev. Lett.}\ }\textbf {\bibinfo {volume} {112}},\ \bibinfo {pages} {120505} (\bibinfo {year} {2014})}\BibitemShut {NoStop}%
\bibitem [{\citenamefont {Roslund}\ \emph {et~al.}(2014)\citenamefont {Roslund}, \citenamefont {de~Araújo}, \citenamefont {Jiang} \emph {et~al.}}]{Roslund:14}%
  \BibitemOpen
  \bibfield  {author} {\bibinfo {author} {\bibfnamefont {J.}~\bibnamefont {Roslund}}, \bibinfo {author} {\bibfnamefont {R.}~\bibnamefont {de~Araújo}}, \bibinfo {author} {\bibfnamefont {S.}~\bibnamefont {Jiang}}, \emph {et~al.},\ }\bibfield  {title} {\bibinfo {title} {Wavelength-multiplexed quantum networks with ultrafast frequency combs},\ }\href {https://doi.org/10.1038/nphoton.2013.340} {\bibfield  {journal} {\bibinfo  {journal} {Nature Photonics}\ }\textbf {\bibinfo {volume} {8}},\ \bibinfo {pages} {109} (\bibinfo {year} {2014})}\BibitemShut {NoStop}%
\bibitem [{\citenamefont {Kouadou}\ \emph {et~al.}(2023)\citenamefont {Kouadou}, \citenamefont {Sansavini}, \citenamefont {Ansquer}, \citenamefont {Henaff}, \citenamefont {Treps},\ and\ \citenamefont {Parigi}}]{Kouadou:23}%
  \BibitemOpen
  \bibfield  {author} {\bibinfo {author} {\bibfnamefont {T.}~\bibnamefont {Kouadou}}, \bibinfo {author} {\bibfnamefont {F.}~\bibnamefont {Sansavini}}, \bibinfo {author} {\bibfnamefont {M.}~\bibnamefont {Ansquer}}, \bibinfo {author} {\bibfnamefont {J.}~\bibnamefont {Henaff}}, \bibinfo {author} {\bibfnamefont {N.}~\bibnamefont {Treps}},\ and\ \bibinfo {author} {\bibfnamefont {V.}~\bibnamefont {Parigi}},\ }\bibfield  {title} {\bibinfo {title} {Spectrally shaped and pulse-by-pulse multiplexed multimode squeezed states of light},\ }\href {https://doi.org/10.1063/5.0156331} {\bibfield  {journal} {\bibinfo  {journal} {APL Photonics}\ }\textbf {\bibinfo {volume} {8}},\ \bibinfo {pages} {086113} (\bibinfo {year} {2023})}\BibitemShut {NoStop}%
\bibitem [{\citenamefont {Zhong}\ \emph {et~al.}(2020)\citenamefont {Zhong}, \citenamefont {Wang}, \citenamefont {Deng}, \citenamefont {Chen}, \citenamefont {Peng}, \citenamefont {Luo}, \citenamefont {Qin}, \citenamefont {Wu}, \citenamefont {Ding}, \citenamefont {Hu}, \citenamefont {Hu}, \citenamefont {Yang}, \citenamefont {Zhang}, \citenamefont {Li}, \citenamefont {Li}, \citenamefont {Jiang}, \citenamefont {Gan}, \citenamefont {Yang}, \citenamefont {You}, \citenamefont {Wang}, \citenamefont {Li}, \citenamefont {Liu}, \citenamefont {Lu},\ and\ \citenamefont {Pan}}]{Zhong:20}%
  \BibitemOpen
  \bibfield  {author} {\bibinfo {author} {\bibfnamefont {H.-S.}\ \bibnamefont {Zhong}}, \bibinfo {author} {\bibfnamefont {H.}~\bibnamefont {Wang}}, \bibinfo {author} {\bibfnamefont {Y.-H.}\ \bibnamefont {Deng}}, \bibinfo {author} {\bibfnamefont {M.-C.}\ \bibnamefont {Chen}}, \bibinfo {author} {\bibfnamefont {L.-C.}\ \bibnamefont {Peng}}, \bibinfo {author} {\bibfnamefont {Y.-H.}\ \bibnamefont {Luo}}, \bibinfo {author} {\bibfnamefont {J.}~\bibnamefont {Qin}}, \bibinfo {author} {\bibfnamefont {D.}~\bibnamefont {Wu}}, \bibinfo {author} {\bibfnamefont {X.}~\bibnamefont {Ding}}, \bibinfo {author} {\bibfnamefont {Y.}~\bibnamefont {Hu}}, \bibinfo {author} {\bibfnamefont {P.}~\bibnamefont {Hu}}, \bibinfo {author} {\bibfnamefont {X.-Y.}\ \bibnamefont {Yang}}, \bibinfo {author} {\bibfnamefont {W.-J.}\ \bibnamefont {Zhang}}, \bibinfo {author} {\bibfnamefont {H.}~\bibnamefont {Li}}, \bibinfo {author} {\bibfnamefont {Y.}~\bibnamefont {Li}}, \bibinfo {author} {\bibfnamefont {X.}~\bibnamefont {Jiang}}, \bibinfo {author}
  {\bibfnamefont {L.}~\bibnamefont {Gan}}, \bibinfo {author} {\bibfnamefont {G.}~\bibnamefont {Yang}}, \bibinfo {author} {\bibfnamefont {L.}~\bibnamefont {You}}, \bibinfo {author} {\bibfnamefont {Z.}~\bibnamefont {Wang}}, \bibinfo {author} {\bibfnamefont {L.}~\bibnamefont {Li}}, \bibinfo {author} {\bibfnamefont {N.-L.}\ \bibnamefont {Liu}}, \bibinfo {author} {\bibfnamefont {C.-Y.}\ \bibnamefont {Lu}},\ and\ \bibinfo {author} {\bibfnamefont {J.-W.}\ \bibnamefont {Pan}},\ }\bibfield  {title} {\bibinfo {title} {Quantum computational advantage using photons},\ }\href {https://doi.org/10.1126/science.abe8770} {\bibfield  {journal} {\bibinfo  {journal} {Science}\ }\textbf {\bibinfo {volume} {370}},\ \bibinfo {pages} {1460} (\bibinfo {year} {2020})}\BibitemShut {NoStop}%
\bibitem [{\citenamefont {Madsen}\ \emph {et~al.}(2022)\citenamefont {Madsen}, \citenamefont {Laudenbach}, \citenamefont {Askarani} \emph {et~al.}}]{Madsen:22}%
  \BibitemOpen
  \bibfield  {author} {\bibinfo {author} {\bibfnamefont {L.~S.}\ \bibnamefont {Madsen}}, \bibinfo {author} {\bibfnamefont {F.}~\bibnamefont {Laudenbach}}, \bibinfo {author} {\bibfnamefont {M.~F.}\ \bibnamefont {Askarani}}, \emph {et~al.},\ }\bibfield  {title} {\bibinfo {title} {Quantum computational advantage with a programmable photonic processor},\ }\href {https://doi.org/10.1038/s41586-022-04725-x} {\bibfield  {journal} {\bibinfo  {journal} {Nature}\ }\textbf {\bibinfo {volume} {606}},\ \bibinfo {pages} {75} (\bibinfo {year} {2022})}\BibitemShut {NoStop}%
\bibitem [{\citenamefont {Pooser}\ and\ \citenamefont {Lawrie}(2015)}]{Pooser:15}%
  \BibitemOpen
  \bibfield  {author} {\bibinfo {author} {\bibfnamefont {R.~C.}\ \bibnamefont {Pooser}}\ and\ \bibinfo {author} {\bibfnamefont {B.}~\bibnamefont {Lawrie}},\ }\bibfield  {title} {\bibinfo {title} {Ultrasensitive measurement of microcantilever displacement below the shot-noise limit},\ }\href {https://doi.org/10.1364/OPTICA.2.000393} {\bibfield  {journal} {\bibinfo  {journal} {Optica}\ }\textbf {\bibinfo {volume} {2}},\ \bibinfo {pages} {393} (\bibinfo {year} {2015})}\BibitemShut {NoStop}%
\bibitem [{\citenamefont {Dowran}\ \emph {et~al.}(2018)\citenamefont {Dowran}, \citenamefont {Kumar}, \citenamefont {Lawrie}, \citenamefont {Pooser},\ and\ \citenamefont {Marino}}]{Dowran:18}%
  \BibitemOpen
  \bibfield  {author} {\bibinfo {author} {\bibfnamefont {M.}~\bibnamefont {Dowran}}, \bibinfo {author} {\bibfnamefont {A.}~\bibnamefont {Kumar}}, \bibinfo {author} {\bibfnamefont {B.~J.}\ \bibnamefont {Lawrie}}, \bibinfo {author} {\bibfnamefont {R.~C.}\ \bibnamefont {Pooser}},\ and\ \bibinfo {author} {\bibfnamefont {A.~M.}\ \bibnamefont {Marino}},\ }\bibfield  {title} {\bibinfo {title} {Quantum-enhanced plasmonic sensing},\ }\href@noop {} {\bibfield  {journal} {\bibinfo  {journal} {Optica}\ }\textbf {\bibinfo {volume} {5}},\ \bibinfo {pages} {628} (\bibinfo {year} {2018})}\BibitemShut {NoStop}%
\bibitem [{\citenamefont {Pooser}\ and\ \citenamefont {Lawrie}(2016)}]{Pooser:16}%
  \BibitemOpen
  \bibfield  {author} {\bibinfo {author} {\bibfnamefont {R.~C.}\ \bibnamefont {Pooser}}\ and\ \bibinfo {author} {\bibfnamefont {B.}~\bibnamefont {Lawrie}},\ }\bibfield  {title} {\bibinfo {title} {Plasmonic trace sensing below the photon shot noise limit},\ }\href {https://doi.org/10.1021/acsphotonics.5b00501} {\bibfield  {journal} {\bibinfo  {journal} {ACS Photonics}\ }\textbf {\bibinfo {volume} {3}},\ \bibinfo {pages} {8} (\bibinfo {year} {2016})}\BibitemShut {NoStop}%
\bibitem [{\citenamefont {Li}\ \emph {et~al.}(2022)\citenamefont {Li}, \citenamefont {Li}, \citenamefont {Liu}, \citenamefont {Yakovlev},\ and\ \citenamefont {Agarwal}}]{Li:22}%
  \BibitemOpen
  \bibfield  {author} {\bibinfo {author} {\bibfnamefont {T.}~\bibnamefont {Li}}, \bibinfo {author} {\bibfnamefont {F.}~\bibnamefont {Li}}, \bibinfo {author} {\bibfnamefont {X.}~\bibnamefont {Liu}}, \bibinfo {author} {\bibfnamefont {V.~V.}\ \bibnamefont {Yakovlev}},\ and\ \bibinfo {author} {\bibfnamefont {G.~S.}\ \bibnamefont {Agarwal}},\ }\bibfield  {title} {\bibinfo {title} {Quantum-enhanced stimulated brillouin scattering spectroscopy and imaging},\ }\href {https://doi.org/10.1364/OPTICA.467635} {\bibfield  {journal} {\bibinfo  {journal} {Optica}\ }\textbf {\bibinfo {volume} {9}},\ \bibinfo {pages} {959} (\bibinfo {year} {2022})}\BibitemShut {NoStop}%
\bibitem [{\citenamefont {Samantaray}\ \emph {et~al.}(2017)\citenamefont {Samantaray}, \citenamefont {Ruo-Berchera}, \citenamefont {Meda} \emph {et~al.}}]{Samantaray:17}%
  \BibitemOpen
  \bibfield  {author} {\bibinfo {author} {\bibfnamefont {N.}~\bibnamefont {Samantaray}}, \bibinfo {author} {\bibfnamefont {I.}~\bibnamefont {Ruo-Berchera}}, \bibinfo {author} {\bibfnamefont {A.}~\bibnamefont {Meda}}, \emph {et~al.},\ }\bibfield  {title} {\bibinfo {title} {Realization of the first sub-shot-noise wide field microscope},\ }\href {https://doi.org/10.1038/lsa.2017.5} {\bibfield  {journal} {\bibinfo  {journal} {Light: Science \& Applications}\ }\textbf {\bibinfo {volume} {6}},\ \bibinfo {pages} {e17005} (\bibinfo {year} {2017})}\BibitemShut {NoStop}%
\bibitem [{\citenamefont {Li}\ \emph {et~al.}(2021)\citenamefont {Li}, \citenamefont {Li}, \citenamefont {Scully},\ and\ \citenamefont {Agarwal}}]{Li:21}%
  \BibitemOpen
  \bibfield  {author} {\bibinfo {author} {\bibfnamefont {F.}~\bibnamefont {Li}}, \bibinfo {author} {\bibfnamefont {T.}~\bibnamefont {Li}}, \bibinfo {author} {\bibfnamefont {M.~O.}\ \bibnamefont {Scully}},\ and\ \bibinfo {author} {\bibfnamefont {G.~S.}\ \bibnamefont {Agarwal}},\ }\bibfield  {title} {\bibinfo {title} {Quantum advantage with seeded squeezed light for absorption measurement},\ }\href {https://doi.org/10.1103/PhysRevApplied.15.044030} {\bibfield  {journal} {\bibinfo  {journal} {Phys. Rev. Appl.}\ }\textbf {\bibinfo {volume} {15}},\ \bibinfo {pages} {044030} (\bibinfo {year} {2021})}\BibitemShut {NoStop}%
\bibitem [{\citenamefont {Li}\ \emph {et~al.}(2024)\citenamefont {Li}, \citenamefont {Cheburkanov}, \citenamefont {Yakovlev}, \citenamefont {Agarwal},\ and\ \citenamefont {Scully}}]{Li:24}%
  \BibitemOpen
  \bibfield  {author} {\bibinfo {author} {\bibfnamefont {T.}~\bibnamefont {Li}}, \bibinfo {author} {\bibfnamefont {V.}~\bibnamefont {Cheburkanov}}, \bibinfo {author} {\bibfnamefont {V.~V.}\ \bibnamefont {Yakovlev}}, \bibinfo {author} {\bibfnamefont {G.~S.}\ \bibnamefont {Agarwal}},\ and\ \bibinfo {author} {\bibfnamefont {M.~O.}\ \bibnamefont {Scully}},\ }\bibfield  {title} {\bibinfo {title} {Harnessing quantum light for microscopic biomechanical imaging of cells and tissues},\ }\href {https://doi.org/10.1073/pnas.2413938121} {\bibfield  {journal} {\bibinfo  {journal} {Proceedings of the National Academy of Sciences}\ }\textbf {\bibinfo {volume} {121}},\ \bibinfo {pages} {e2413938121} (\bibinfo {year} {2024})}\BibitemShut {NoStop}%
\bibitem [{\citenamefont {Casacio}\ \emph {et~al.}(2021)\citenamefont {Casacio}, \citenamefont {Madsen}, \citenamefont {Terrasson} \emph {et~al.}}]{Casacio:21}%
  \BibitemOpen
  \bibfield  {author} {\bibinfo {author} {\bibfnamefont {C.~A.}\ \bibnamefont {Casacio}}, \bibinfo {author} {\bibfnamefont {L.~S.}\ \bibnamefont {Madsen}}, \bibinfo {author} {\bibfnamefont {A.}~\bibnamefont {Terrasson}}, \emph {et~al.},\ }\bibfield  {title} {\bibinfo {title} {Quantum-enhanced nonlinear microscopy},\ }\href {https://doi.org/10.1038/s41586-021-03528-w} {\bibfield  {journal} {\bibinfo  {journal} {Nature}\ }\textbf {\bibinfo {volume} {594}},\ \bibinfo {pages} {201} (\bibinfo {year} {2021})}\BibitemShut {NoStop}%
\bibitem [{\citenamefont {Fürst}\ \emph {et~al.}(2011)\citenamefont {Fürst}, \citenamefont {Strekalov}, \citenamefont {Elser}, \citenamefont {Aiello}, \citenamefont {Andersen}, \citenamefont {Marquardt},\ and\ \citenamefont {Leuchs}}]{Furst:11}%
  \BibitemOpen
  \bibfield  {author} {\bibinfo {author} {\bibfnamefont {J.~U.}\ \bibnamefont {Fürst}}, \bibinfo {author} {\bibfnamefont {D.~V.}\ \bibnamefont {Strekalov}}, \bibinfo {author} {\bibfnamefont {D.}~\bibnamefont {Elser}}, \bibinfo {author} {\bibfnamefont {A.}~\bibnamefont {Aiello}}, \bibinfo {author} {\bibfnamefont {U.~L.}\ \bibnamefont {Andersen}}, \bibinfo {author} {\bibfnamefont {C.}~\bibnamefont {Marquardt}},\ and\ \bibinfo {author} {\bibfnamefont {G.}~\bibnamefont {Leuchs}},\ }\bibfield  {title} {\bibinfo {title} {Quantum light from a whispering-gallery-mode disk resonator},\ }\href@noop {} {\bibfield  {journal} {\bibinfo  {journal} {Phys. Rev. Lett.}\ }\textbf {\bibinfo {volume} {106}},\ \bibinfo {pages} {113901} (\bibinfo {year} {2011})}\BibitemShut {NoStop}%
\bibitem [{\citenamefont {Dutt}\ \emph {et~al.}(2015)\citenamefont {Dutt}, \citenamefont {Luke}, \citenamefont {Manipatruni}, \citenamefont {Gaeta}, \citenamefont {Nussenzveig},\ and\ \citenamefont {Lipson}}]{Dutt:15}%
  \BibitemOpen
  \bibfield  {author} {\bibinfo {author} {\bibfnamefont {A.}~\bibnamefont {Dutt}}, \bibinfo {author} {\bibfnamefont {K.}~\bibnamefont {Luke}}, \bibinfo {author} {\bibfnamefont {S.}~\bibnamefont {Manipatruni}}, \bibinfo {author} {\bibfnamefont {A.~L.}\ \bibnamefont {Gaeta}}, \bibinfo {author} {\bibfnamefont {P.}~\bibnamefont {Nussenzveig}},\ and\ \bibinfo {author} {\bibfnamefont {M.}~\bibnamefont {Lipson}},\ }\bibfield  {title} {\bibinfo {title} {On-chip optical squeezing},\ }\href {http://link.aps.org/doi/10.1103/PhysRevApplied.3.044005} {\bibfield  {journal} {\bibinfo  {journal} {Phys. Rev. Applied}\ }\textbf {\bibinfo {volume} {3}},\ \bibinfo {pages} {044005} (\bibinfo {year} {2015})}\BibitemShut {NoStop}%
\bibitem [{\citenamefont {Dutt}\ \emph {et~al.}(2016)\citenamefont {Dutt}, \citenamefont {Miller}, \citenamefont {Luke}, \citenamefont {Cardenas}, \citenamefont {Gaeta}, \citenamefont {Nussenzveig},\ and\ \citenamefont {Lipson}}]{Dutt:16}%
  \BibitemOpen
  \bibfield  {author} {\bibinfo {author} {\bibfnamefont {A.}~\bibnamefont {Dutt}}, \bibinfo {author} {\bibfnamefont {S.}~\bibnamefont {Miller}}, \bibinfo {author} {\bibfnamefont {K.}~\bibnamefont {Luke}}, \bibinfo {author} {\bibfnamefont {J.}~\bibnamefont {Cardenas}}, \bibinfo {author} {\bibfnamefont {A.~L.}\ \bibnamefont {Gaeta}}, \bibinfo {author} {\bibfnamefont {P.}~\bibnamefont {Nussenzveig}},\ and\ \bibinfo {author} {\bibfnamefont {M.}~\bibnamefont {Lipson}},\ }\bibfield  {title} {\bibinfo {title} {Tunable squeezing using coupled ring resonators on a silicon nitride chip},\ }\href {https://doi.org/10.1364/OL.41.000223} {\bibfield  {journal} {\bibinfo  {journal} {Opt. Lett.}\ }\textbf {\bibinfo {volume} {41}},\ \bibinfo {pages} {223} (\bibinfo {year} {2016})}\BibitemShut {NoStop}%
\bibitem [{\citenamefont {Kögler}\ \emph {et~al.}(2024)\citenamefont {Kögler}, \citenamefont {Rickli}, \citenamefont {Domeneguetti}, \citenamefont {Ji}, \citenamefont {Gaeta}, \citenamefont {Lipson}, \citenamefont {Martinelli},\ and\ \citenamefont {Nussenzveig}}]{Kogler:24}%
  \BibitemOpen
  \bibfield  {author} {\bibinfo {author} {\bibfnamefont {R.~A.}\ \bibnamefont {Kögler}}, \bibinfo {author} {\bibfnamefont {G.~C.}\ \bibnamefont {Rickli}}, \bibinfo {author} {\bibfnamefont {R.~R.}\ \bibnamefont {Domeneguetti}}, \bibinfo {author} {\bibfnamefont {X.}~\bibnamefont {Ji}}, \bibinfo {author} {\bibfnamefont {A.~L.}\ \bibnamefont {Gaeta}}, \bibinfo {author} {\bibfnamefont {M.}~\bibnamefont {Lipson}}, \bibinfo {author} {\bibfnamefont {M.}~\bibnamefont {Martinelli}},\ and\ \bibinfo {author} {\bibfnamefont {P.}~\bibnamefont {Nussenzveig}},\ }\bibfield  {title} {\bibinfo {title} {Quantum state tomography in a third-order integrated optical parametric oscillator},\ }\href@noop {} {\bibfield  {journal} {\bibinfo  {journal} {Opt. Lett.}\ }\textbf {\bibinfo {volume} {49}},\ \bibinfo {pages} {3150} (\bibinfo {year} {2024})}\BibitemShut {NoStop}%
\bibitem [{\citenamefont {Wang}\ \emph {et~al.}(2024)\citenamefont {Wang}, \citenamefont {Li}, \citenamefont {Wang}, \citenamefont {Zhou}, \citenamefont {Cheng}, \citenamefont {Jing}, \citenamefont {Sun}, \citenamefont {Li}, \citenamefont {Li}, \citenamefont {Gong}, \citenamefont {He}, \citenamefont {Li},\ and\ \citenamefont {Yang}}]{Ze:24}%
  \BibitemOpen
  \bibfield  {author} {\bibinfo {author} {\bibfnamefont {Z.}~\bibnamefont {Wang}}, \bibinfo {author} {\bibfnamefont {K.}~\bibnamefont {Li}}, \bibinfo {author} {\bibfnamefont {Y.}~\bibnamefont {Wang}}, \bibinfo {author} {\bibfnamefont {X.}~\bibnamefont {Zhou}}, \bibinfo {author} {\bibfnamefont {Y.}~\bibnamefont {Cheng}}, \bibinfo {author} {\bibfnamefont {B.}~\bibnamefont {Jing}}, \bibinfo {author} {\bibfnamefont {F.}~\bibnamefont {Sun}}, \bibinfo {author} {\bibfnamefont {J.}~\bibnamefont {Li}}, \bibinfo {author} {\bibfnamefont {Z.}~\bibnamefont {Li}}, \bibinfo {author} {\bibfnamefont {Q.}~\bibnamefont {Gong}}, \bibinfo {author} {\bibfnamefont {Q.}~\bibnamefont {He}}, \bibinfo {author} {\bibfnamefont {B.-B.}\ \bibnamefont {Li}},\ and\ \bibinfo {author} {\bibfnamefont {Q.-F.}\ \bibnamefont {Yang}},\ }\href {https://doi.org/10.48550/arXiv.2406.10715} {\bibinfo {title} {Chip-scale generation of 60-mode continuous-variable cluster states}} (\bibinfo {year} {2024}),\ \Eprint {https://arxiv.org/abs/2406.10715}
  {arXiv:2406.10715} \BibitemShut {NoStop}%
\bibitem [{\citenamefont {Yang}\ \emph {et~al.}(2021)\citenamefont {Yang}, \citenamefont {Jahanbozorgi}, \citenamefont {Jeong}, \citenamefont {Sun}, \citenamefont {Pfister}, \citenamefont {Lee},\ and\ \citenamefont {Yi}}]{Yang:21}%
  \BibitemOpen
  \bibfield  {author} {\bibinfo {author} {\bibfnamefont {Z.}~\bibnamefont {Yang}}, \bibinfo {author} {\bibfnamefont {M.}~\bibnamefont {Jahanbozorgi}}, \bibinfo {author} {\bibfnamefont {D.}~\bibnamefont {Jeong}}, \bibinfo {author} {\bibfnamefont {S.}~\bibnamefont {Sun}}, \bibinfo {author} {\bibfnamefont {O.}~\bibnamefont {Pfister}}, \bibinfo {author} {\bibfnamefont {H.}~\bibnamefont {Lee}},\ and\ \bibinfo {author} {\bibfnamefont {X.}~\bibnamefont {Yi}},\ }\bibfield  {title} {\bibinfo {title} {A squeezed quantum microcomb on a chip},\ }\href@noop {} {\bibfield  {journal} {\bibinfo  {journal} {Nature Communications}\ }\textbf {\bibinfo {volume} {12}},\ \bibinfo {pages} {4781} (\bibinfo {year} {2021})}\BibitemShut {NoStop}%
\bibitem [{\citenamefont {Zhao}\ \emph {et~al.}(2020)\citenamefont {Zhao}, \citenamefont {Okawachi}, \citenamefont {Jang}, \citenamefont {Ji}, \citenamefont {Lipson},\ and\ \citenamefont {Gaeta}}]{Zhao:20}%
  \BibitemOpen
  \bibfield  {author} {\bibinfo {author} {\bibfnamefont {Y.}~\bibnamefont {Zhao}}, \bibinfo {author} {\bibfnamefont {Y.}~\bibnamefont {Okawachi}}, \bibinfo {author} {\bibfnamefont {J.~K.}\ \bibnamefont {Jang}}, \bibinfo {author} {\bibfnamefont {X.}~\bibnamefont {Ji}}, \bibinfo {author} {\bibfnamefont {M.}~\bibnamefont {Lipson}},\ and\ \bibinfo {author} {\bibfnamefont {A.~L.}\ \bibnamefont {Gaeta}},\ }\bibfield  {title} {\bibinfo {title} {Near-degenerate quadrature-squeezed vacuum generation on a silicon-nitride chip},\ }\href {https://doi.org/10.1103/PhysRevLett.124.193601} {\bibfield  {journal} {\bibinfo  {journal} {Phys. Rev. Lett.}\ }\textbf {\bibinfo {volume} {124}},\ \bibinfo {pages} {193601} (\bibinfo {year} {2020})}\BibitemShut {NoStop}%
\bibitem [{\citenamefont {Chen}\ \emph {et~al.}(2022)\citenamefont {Chen}, \citenamefont {Briggs}, \citenamefont {Hou},\ and\ \citenamefont {Fan}}]{Chen:22}%
  \BibitemOpen
  \bibfield  {author} {\bibinfo {author} {\bibfnamefont {P.-K.}\ \bibnamefont {Chen}}, \bibinfo {author} {\bibfnamefont {I.}~\bibnamefont {Briggs}}, \bibinfo {author} {\bibfnamefont {S.}~\bibnamefont {Hou}},\ and\ \bibinfo {author} {\bibfnamefont {L.}~\bibnamefont {Fan}},\ }\bibfield  {title} {\bibinfo {title} {Ultra-broadband quadrature squeezing with thin-film lithium niobate nanophotonics},\ }\href {https://doi.org/10.1364/OL.447695} {\bibfield  {journal} {\bibinfo  {journal} {Opt. Lett.}\ }\textbf {\bibinfo {volume} {47}},\ \bibinfo {pages} {1506} (\bibinfo {year} {2022})}\BibitemShut {NoStop}%
\bibitem [{\citenamefont {Jahanbozorgi}\ \emph {et~al.}(2023)\citenamefont {Jahanbozorgi}, \citenamefont {Yang}, \citenamefont {Sun}, \citenamefont {Chen}, \citenamefont {Liu}, \citenamefont {Wang},\ and\ \citenamefont {Yi}}]{Jahanbozorgi:23}%
  \BibitemOpen
  \bibfield  {author} {\bibinfo {author} {\bibfnamefont {M.}~\bibnamefont {Jahanbozorgi}}, \bibinfo {author} {\bibfnamefont {Z.}~\bibnamefont {Yang}}, \bibinfo {author} {\bibfnamefont {S.}~\bibnamefont {Sun}}, \bibinfo {author} {\bibfnamefont {H.}~\bibnamefont {Chen}}, \bibinfo {author} {\bibfnamefont {R.}~\bibnamefont {Liu}}, \bibinfo {author} {\bibfnamefont {B.}~\bibnamefont {Wang}},\ and\ \bibinfo {author} {\bibfnamefont {X.}~\bibnamefont {Yi}},\ }\bibfield  {title} {\bibinfo {title} {Generation of squeezed quantum microcombs with silicon nitride integrated photonic circuits},\ }\href@noop {} {\bibfield  {journal} {\bibinfo  {journal} {Optica}\ }\textbf {\bibinfo {volume} {10}},\ \bibinfo {pages} {1100} (\bibinfo {year} {2023})}\BibitemShut {NoStop}%
\bibitem [{\citenamefont {Cernansky}\ and\ \citenamefont {Politi}(2020)}]{Cernansky:20}%
  \BibitemOpen
  \bibfield  {author} {\bibinfo {author} {\bibfnamefont {R.}~\bibnamefont {Cernansky}}\ and\ \bibinfo {author} {\bibfnamefont {A.}~\bibnamefont {Politi}},\ }\bibfield  {title} {\bibinfo {title} {Nanophotonic source of quadrature squeezing via self-phase modulation},\ }\href@noop {} {\bibfield  {journal} {\bibinfo  {journal} {APL Photonics}\ }\textbf {\bibinfo {volume} {5}},\ \bibinfo {pages} {101303} (\bibinfo {year} {2020})}\BibitemShut {NoStop}%
\bibitem [{\citenamefont {Vaidya}\ \emph {et~al.}(2020)\citenamefont {Vaidya}, \citenamefont {Morrison}, \citenamefont {Helt}, \citenamefont {Shahrokshahi}, \citenamefont {Mahler}, \citenamefont {Collins}, \citenamefont {Tan}, \citenamefont {Lavoie}, \citenamefont {Repingon}, \citenamefont {Menotti}, \citenamefont {Quesada}, \citenamefont {Pooser}, \citenamefont {Lita}, \citenamefont {Gerrits}, \citenamefont {Nam},\ and\ \citenamefont {Vernon}}]{Vaidya:20}%
  \BibitemOpen
  \bibfield  {author} {\bibinfo {author} {\bibfnamefont {V.~D.}\ \bibnamefont {Vaidya}}, \bibinfo {author} {\bibfnamefont {B.}~\bibnamefont {Morrison}}, \bibinfo {author} {\bibfnamefont {L.~G.}\ \bibnamefont {Helt}}, \bibinfo {author} {\bibfnamefont {R.}~\bibnamefont {Shahrokshahi}}, \bibinfo {author} {\bibfnamefont {D.~H.}\ \bibnamefont {Mahler}}, \bibinfo {author} {\bibfnamefont {M.~J.}\ \bibnamefont {Collins}}, \bibinfo {author} {\bibfnamefont {K.}~\bibnamefont {Tan}}, \bibinfo {author} {\bibfnamefont {J.}~\bibnamefont {Lavoie}}, \bibinfo {author} {\bibfnamefont {A.}~\bibnamefont {Repingon}}, \bibinfo {author} {\bibfnamefont {M.}~\bibnamefont {Menotti}}, \bibinfo {author} {\bibfnamefont {N.}~\bibnamefont {Quesada}}, \bibinfo {author} {\bibfnamefont {R.~C.}\ \bibnamefont {Pooser}}, \bibinfo {author} {\bibfnamefont {A.~E.}\ \bibnamefont {Lita}}, \bibinfo {author} {\bibfnamefont {T.}~\bibnamefont {Gerrits}}, \bibinfo {author} {\bibfnamefont {S.~W.}\ \bibnamefont {Nam}},\ and\ \bibinfo {author} {\bibfnamefont
  {Z.}~\bibnamefont {Vernon}},\ }\bibfield  {title} {\bibinfo {title} {Broadband quadrature-squeezed vacuum and nonclassical photon number correlations from a nanophotonic device},\ }\href {https://doi.org/10.1126/sciadv.aba9186} {\bibfield  {journal} {\bibinfo  {journal} {Science Advances}\ }\textbf {\bibinfo {volume} {6}},\ \bibinfo {pages} {eaba9186} (\bibinfo {year} {2020})}\BibitemShut {NoStop}%
\bibitem [{\citenamefont {Stokowski}\ \emph {et~al.}(2023)\citenamefont {Stokowski}, \citenamefont {McKenna}, \citenamefont {Park} \emph {et~al.}}]{Stokowski:23}%
  \BibitemOpen
  \bibfield  {author} {\bibinfo {author} {\bibfnamefont {H.~S.}\ \bibnamefont {Stokowski}}, \bibinfo {author} {\bibfnamefont {T.~P.}\ \bibnamefont {McKenna}}, \bibinfo {author} {\bibfnamefont {T.}~\bibnamefont {Park}}, \emph {et~al.},\ }\bibfield  {title} {\bibinfo {title} {Integrated quantum optical phase sensor in thin film lithium niobate},\ }\href {https://doi.org/10.1038/s41467-023-38246-6} {\bibfield  {journal} {\bibinfo  {journal} {Nature Communications}\ }\textbf {\bibinfo {volume} {14}},\ \bibinfo {pages} {3355} (\bibinfo {year} {2023})}\BibitemShut {NoStop}%
\bibitem [{\citenamefont {Park}\ \emph {et~al.}(2024)\citenamefont {Park}, \citenamefont {Stokowski}, \citenamefont {Ansari}, \citenamefont {Gyger}, \citenamefont {Multani}, \citenamefont {Celik}, \citenamefont {Hwang}, \citenamefont {Dean}, \citenamefont {Mayor}, \citenamefont {McKenna}, \citenamefont {Fejer},\ and\ \citenamefont {Safavi-Naeini}}]{Park:24}%
  \BibitemOpen
  \bibfield  {author} {\bibinfo {author} {\bibfnamefont {T.}~\bibnamefont {Park}}, \bibinfo {author} {\bibfnamefont {H.}~\bibnamefont {Stokowski}}, \bibinfo {author} {\bibfnamefont {V.}~\bibnamefont {Ansari}}, \bibinfo {author} {\bibfnamefont {S.}~\bibnamefont {Gyger}}, \bibinfo {author} {\bibfnamefont {K.~K.~S.}\ \bibnamefont {Multani}}, \bibinfo {author} {\bibfnamefont {O.~T.}\ \bibnamefont {Celik}}, \bibinfo {author} {\bibfnamefont {A.~Y.}\ \bibnamefont {Hwang}}, \bibinfo {author} {\bibfnamefont {D.~J.}\ \bibnamefont {Dean}}, \bibinfo {author} {\bibfnamefont {F.}~\bibnamefont {Mayor}}, \bibinfo {author} {\bibfnamefont {T.~P.}\ \bibnamefont {McKenna}}, \bibinfo {author} {\bibfnamefont {M.~M.}\ \bibnamefont {Fejer}},\ and\ \bibinfo {author} {\bibfnamefont {A.}~\bibnamefont {Safavi-Naeini}},\ }\bibfield  {title} {\bibinfo {title} {Single-mode squeezed-light generation and tomography with an integrated optical parametric oscillator},\ }\href@noop {} {\bibfield  {journal} {\bibinfo  {journal} {Sci. Adv.}\
  }\textbf {\bibinfo {volume} {10}},\ \bibinfo {pages} {eadl1814} (\bibinfo {year} {2024})}\BibitemShut {NoStop}%
\bibitem [{\citenamefont {Takanashi}\ \emph {et~al.}(2020)\citenamefont {Takanashi}, \citenamefont {Inoue}, \citenamefont {Kashiwazaki}, \citenamefont {Kazama}, \citenamefont {Enbutsu}, \citenamefont {Kasahara}, \citenamefont {Umeki},\ and\ \citenamefont {Furusawa}}]{Takanashi:20}%
  \BibitemOpen
  \bibfield  {author} {\bibinfo {author} {\bibfnamefont {N.}~\bibnamefont {Takanashi}}, \bibinfo {author} {\bibfnamefont {A.}~\bibnamefont {Inoue}}, \bibinfo {author} {\bibfnamefont {T.}~\bibnamefont {Kashiwazaki}}, \bibinfo {author} {\bibfnamefont {T.}~\bibnamefont {Kazama}}, \bibinfo {author} {\bibfnamefont {K.}~\bibnamefont {Enbutsu}}, \bibinfo {author} {\bibfnamefont {R.}~\bibnamefont {Kasahara}}, \bibinfo {author} {\bibfnamefont {T.}~\bibnamefont {Umeki}},\ and\ \bibinfo {author} {\bibfnamefont {A.}~\bibnamefont {Furusawa}},\ }\bibfield  {title} {\bibinfo {title} {All-optical phase-sensitive detection for ultra-fast quantum computation},\ }\href {https://doi.org/10.1364/OE.405832} {\bibfield  {journal} {\bibinfo  {journal} {Opt. Express}\ }\textbf {\bibinfo {volume} {28}},\ \bibinfo {pages} {34916} (\bibinfo {year} {2020})}\BibitemShut {NoStop}%
\bibitem [{\citenamefont {Nehra}\ \emph {et~al.}(2022)\citenamefont {Nehra}, \citenamefont {Sekine}, \citenamefont {Ledezma}, \citenamefont {Guo}, \citenamefont {Gray}, \citenamefont {Roy},\ and\ \citenamefont {Marandi}}]{Nehra:22}%
  \BibitemOpen
  \bibfield  {author} {\bibinfo {author} {\bibfnamefont {R.}~\bibnamefont {Nehra}}, \bibinfo {author} {\bibfnamefont {R.}~\bibnamefont {Sekine}}, \bibinfo {author} {\bibfnamefont {L.}~\bibnamefont {Ledezma}}, \bibinfo {author} {\bibfnamefont {Q.}~\bibnamefont {Guo}}, \bibinfo {author} {\bibfnamefont {R.~M.}\ \bibnamefont {Gray}}, \bibinfo {author} {\bibfnamefont {A.}~\bibnamefont {Roy}},\ and\ \bibinfo {author} {\bibfnamefont {A.}~\bibnamefont {Marandi}},\ }\bibfield  {title} {\bibinfo {title} {Few-cycle vacuum squeezing in nanophotonics},\ }\href@noop {} {\bibfield  {journal} {\bibinfo  {journal} {Science}\ }\textbf {\bibinfo {volume} {377}},\ \bibinfo {pages} {1333} (\bibinfo {year} {2022})}\BibitemShut {NoStop}%
\bibitem [{\citenamefont {Brusaschi}\ \emph {et~al.}(2024)\citenamefont {Brusaschi}, \citenamefont {Borghi}, \citenamefont {Bacchi}, \citenamefont {Liscidini}, \citenamefont {Galli},\ and\ \citenamefont {Bajoni}}]{Brusaschi:24}%
  \BibitemOpen
  \bibfield  {author} {\bibinfo {author} {\bibfnamefont {E.}~\bibnamefont {Brusaschi}}, \bibinfo {author} {\bibfnamefont {M.}~\bibnamefont {Borghi}}, \bibinfo {author} {\bibfnamefont {M.}~\bibnamefont {Bacchi}}, \bibinfo {author} {\bibfnamefont {M.}~\bibnamefont {Liscidini}}, \bibinfo {author} {\bibfnamefont {M.}~\bibnamefont {Galli}},\ and\ \bibinfo {author} {\bibfnamefont {D.}~\bibnamefont {Bajoni}},\ }\bibfield  {title} {\bibinfo {title} {Photon number distribution of squeezed light from a silicon nitride microresonator measured without photon number resolving detectors},\ }\href@noop {} {\bibfield  {journal} {\bibinfo  {journal} {Optica Quantum}\ }\textbf {\bibinfo {volume} {2}},\ \bibinfo {pages} {214} (\bibinfo {year} {2024})}\BibitemShut {NoStop}%
\bibitem [{\citenamefont {Fabre}\ \emph {et~al.}(1989)\citenamefont {Fabre}, \citenamefont {Giacobino}, \citenamefont {Heidmann},\ and\ \citenamefont {Reynaud}}]{Fabre:89}%
  \BibitemOpen
  \bibfield  {author} {\bibinfo {author} {\bibfnamefont {C.}~\bibnamefont {Fabre}}, \bibinfo {author} {\bibfnamefont {E.}~\bibnamefont {Giacobino}}, \bibinfo {author} {\bibfnamefont {A.}~\bibnamefont {Heidmann}},\ and\ \bibinfo {author} {\bibfnamefont {S.}~\bibnamefont {Reynaud}},\ }\bibfield  {title} {\bibinfo {title} {Noise characteristics of a non-degenerate optical parametric oscillator - application to quantum noise reduction},\ }\href@noop {} {\bibfield  {journal} {\bibinfo  {journal} {J. Phys. France}\ }\textbf {\bibinfo {volume} {50}},\ \bibinfo {pages} {1209} (\bibinfo {year} {1989})}\BibitemShut {NoStop}%
\bibitem [{\citenamefont {Chembo}(2016)}]{Chembo:16}%
  \BibitemOpen
  \bibfield  {author} {\bibinfo {author} {\bibfnamefont {Y.~K.}\ \bibnamefont {Chembo}},\ }\bibfield  {title} {\bibinfo {title} {Quantum dynamics of kerr optical frequency combs below and above threshold: Spontaneous four-wave mixing, entanglement, and squeezed states of light},\ }\href {https://doi.org/10.1103/PhysRevA.93.033820} {\bibfield  {journal} {\bibinfo  {journal} {Phys. Rev. A}\ }\textbf {\bibinfo {volume} {93}},\ \bibinfo {pages} {033820} (\bibinfo {year} {2016})}\BibitemShut {NoStop}%
\bibitem [{\citenamefont {Ji}\ \emph {et~al.}(2017)\citenamefont {Ji}, \citenamefont {Barbosa}, \citenamefont {Roberts}, \citenamefont {Dutt}, \citenamefont {Cardenas}, \citenamefont {Okawachi}, \citenamefont {Bryant}, \citenamefont {Gaeta},\ and\ \citenamefont {Lipson}}]{Ji:17}%
  \BibitemOpen
  \bibfield  {author} {\bibinfo {author} {\bibfnamefont {X.}~\bibnamefont {Ji}}, \bibinfo {author} {\bibfnamefont {F.~A.~S.}\ \bibnamefont {Barbosa}}, \bibinfo {author} {\bibfnamefont {S.~P.}\ \bibnamefont {Roberts}}, \bibinfo {author} {\bibfnamefont {A.}~\bibnamefont {Dutt}}, \bibinfo {author} {\bibfnamefont {J.}~\bibnamefont {Cardenas}}, \bibinfo {author} {\bibfnamefont {Y.}~\bibnamefont {Okawachi}}, \bibinfo {author} {\bibfnamefont {A.}~\bibnamefont {Bryant}}, \bibinfo {author} {\bibfnamefont {A.~L.}\ \bibnamefont {Gaeta}},\ and\ \bibinfo {author} {\bibfnamefont {M.}~\bibnamefont {Lipson}},\ }\bibfield  {title} {\bibinfo {title} {Ultra-low-loss on-chip resonators with sub-milliwatt parametric oscillation threshold},\ }\href@noop {} {\bibfield  {journal} {\bibinfo  {journal} {Optica}\ }\textbf {\bibinfo {volume} {4}},\ \bibinfo {pages} {619} (\bibinfo {year} {2017})}\BibitemShut {NoStop}%
\bibitem [{\citenamefont {Dutt}\ \emph {et~al.}(2024)\citenamefont {Dutt}, \citenamefont {Mohanty}, \citenamefont {Gaeta},\ and\ \citenamefont {Lipson}}]{Dutt:24}%
  \BibitemOpen
  \bibfield  {author} {\bibinfo {author} {\bibfnamefont {A.}~\bibnamefont {Dutt}}, \bibinfo {author} {\bibfnamefont {A.}~\bibnamefont {Mohanty}}, \bibinfo {author} {\bibfnamefont {A.~L.}\ \bibnamefont {Gaeta}},\ and\ \bibinfo {author} {\bibfnamefont {M.}~\bibnamefont {Lipson}},\ }\bibfield  {title} {\bibinfo {title} {Nonlinear and quantum photonics using integrated optical materials},\ }\href {https://doi.org/10.1038/s41578-024-00668-z} {\bibfield  {journal} {\bibinfo  {journal} {Nat. Rev. Mater.}\ }\textbf {\bibinfo {volume} {9}},\ \bibinfo {pages} {321} (\bibinfo {year} {2024})}\BibitemShut {NoStop}%
\bibitem [{\citenamefont {Hausmann}\ \emph {et~al.}(2014)\citenamefont {Hausmann}, \citenamefont {Bulu}, \citenamefont {Venkataraman}, \citenamefont {Deotare},\ and\ \citenamefont {Lončar}}]{hausmann_diamond_2014}%
  \BibitemOpen
  \bibfield  {author} {\bibinfo {author} {\bibfnamefont {B.~J.~M.}\ \bibnamefont {Hausmann}}, \bibinfo {author} {\bibfnamefont {I.}~\bibnamefont {Bulu}}, \bibinfo {author} {\bibfnamefont {V.}~\bibnamefont {Venkataraman}}, \bibinfo {author} {\bibfnamefont {P.}~\bibnamefont {Deotare}},\ and\ \bibinfo {author} {\bibfnamefont {M.}~\bibnamefont {Lončar}},\ }\bibfield  {title} {\bibinfo {title} {Diamond nonlinear photonics},\ }\href {https://doi.org/10.1038/nphoton.2014.72} {\bibfield  {journal} {\bibinfo  {journal} {Nat Photon}\ }\textbf {\bibinfo {volume} {8}},\ \bibinfo {pages} {369} (\bibinfo {year} {2014})}\BibitemShut {NoStop}%
\bibitem [{\citenamefont {Herr}\ \emph {et~al.}(2012)\citenamefont {Herr}, \citenamefont {Hartinger}, \citenamefont {Riemensberger} \emph {et~al.}}]{Herr:12}%
  \BibitemOpen
  \bibfield  {author} {\bibinfo {author} {\bibfnamefont {T.}~\bibnamefont {Herr}}, \bibinfo {author} {\bibfnamefont {K.}~\bibnamefont {Hartinger}}, \bibinfo {author} {\bibfnamefont {J.}~\bibnamefont {Riemensberger}}, \emph {et~al.},\ }\bibfield  {title} {\bibinfo {title} {Universal formation dynamics and noise of kerr-frequency combs in microresonators},\ }\href {https://doi.org/10.1038/nphoton.2012.127} {\bibfield  {journal} {\bibinfo  {journal} {Nature Photonics}\ }\textbf {\bibinfo {volume} {6}},\ \bibinfo {pages} {480} (\bibinfo {year} {2012})}\BibitemShut {NoStop}%
\bibitem [{\citenamefont {Huffman}\ \emph {et~al.}(2017)\citenamefont {Huffman}, \citenamefont {Baney},\ and\ \citenamefont {Blumenthal}}]{Huffman:17}%
  \BibitemOpen
  \bibfield  {author} {\bibinfo {author} {\bibfnamefont {T.}~\bibnamefont {Huffman}}, \bibinfo {author} {\bibfnamefont {D.}~\bibnamefont {Baney}},\ and\ \bibinfo {author} {\bibfnamefont {D.~J.}\ \bibnamefont {Blumenthal}},\ }\href {https://arxiv.org/abs/1708.06344} {\bibinfo {title} {High extinction ratio widely tunable low-loss integrated si3n4 third-order filter}} (\bibinfo {year} {2017}),\ \Eprint {https://arxiv.org/abs/1708.06344} {arXiv:1708.06344 [physics.app-ph]} \BibitemShut {NoStop}%
\bibitem [{\citenamefont {Nie}\ \emph {et~al.}(2019)\citenamefont {Nie}, \citenamefont {Turk}, \citenamefont {Li}, \citenamefont {Liu},\ and\ \citenamefont {Baets}}]{Nie:19}%
  \BibitemOpen
  \bibfield  {author} {\bibinfo {author} {\bibfnamefont {X.}~\bibnamefont {Nie}}, \bibinfo {author} {\bibfnamefont {N.}~\bibnamefont {Turk}}, \bibinfo {author} {\bibfnamefont {Y.}~\bibnamefont {Li}}, \bibinfo {author} {\bibfnamefont {Z.}~\bibnamefont {Liu}},\ and\ \bibinfo {author} {\bibfnamefont {R.}~\bibnamefont {Baets}},\ }\bibfield  {title} {\bibinfo {title} {High extinction ratio on-chip pump-rejection filter based on cascaded grating-assisted contra-directional couplers in silicon nitride rib waveguides},\ }\href {https://doi.org/10.1364/OL.44.002310} {\bibfield  {journal} {\bibinfo  {journal} {Opt. Lett.}\ }\textbf {\bibinfo {volume} {44}},\ \bibinfo {pages} {2310} (\bibinfo {year} {2019})}\BibitemShut {NoStop}%
\bibitem [{\citenamefont {Afifi}\ \emph {et~al.}(2021)\citenamefont {Afifi}, \citenamefont {Hammood}, \citenamefont {Jaeger}, \citenamefont {Shekhar}, \citenamefont {Chrostowski},\ and\ \citenamefont {Young}}]{Afifi:21}%
  \BibitemOpen
  \bibfield  {author} {\bibinfo {author} {\bibfnamefont {A.~E.}\ \bibnamefont {Afifi}}, \bibinfo {author} {\bibfnamefont {M.}~\bibnamefont {Hammood}}, \bibinfo {author} {\bibfnamefont {N.~A.~F.}\ \bibnamefont {Jaeger}}, \bibinfo {author} {\bibfnamefont {S.}~\bibnamefont {Shekhar}}, \bibinfo {author} {\bibfnamefont {L.}~\bibnamefont {Chrostowski}},\ and\ \bibinfo {author} {\bibfnamefont {J.~F.}\ \bibnamefont {Young}},\ }\bibfield  {title} {\bibinfo {title} {Contra-directional pump reject filters integrated with a micro-ring resonator photon-pair source in silicon},\ }\href {https://doi.org/10.1364/OE.431921} {\bibfield  {journal} {\bibinfo  {journal} {Opt. Express}\ }\textbf {\bibinfo {volume} {29}},\ \bibinfo {pages} {25173} (\bibinfo {year} {2021})}\BibitemShut {NoStop}%
\bibitem [{\citenamefont {Singh}\ \emph {et~al.}(2022)\citenamefont {Singh}, \citenamefont {Belansky},\ and\ \citenamefont {Soltani}}]{Singh:22}%
  \BibitemOpen
  \bibfield  {author} {\bibinfo {author} {\bibfnamefont {A.}~\bibnamefont {Singh}}, \bibinfo {author} {\bibfnamefont {R.}~\bibnamefont {Belansky}},\ and\ \bibinfo {author} {\bibfnamefont {M.}~\bibnamefont {Soltani}},\ }\bibfield  {title} {\bibinfo {title} {Ultraflat bandpass, high extinction, and tunable silicon photonic filters},\ }\href {https://doi.org/10.1364/OE.469864} {\bibfield  {journal} {\bibinfo  {journal} {Opt. Express}\ }\textbf {\bibinfo {volume} {30}},\ \bibinfo {pages} {43787} (\bibinfo {year} {2022})}\BibitemShut {NoStop}%
\bibitem [{\citenamefont {Gao}\ \emph {et~al.}(2023)\citenamefont {Gao}, \citenamefont {Tzu}, \citenamefont {Fatema}, \citenamefont {Guo}, \citenamefont {Yu}, \citenamefont {Navickaite}, \citenamefont {Zervas}, \citenamefont {Geiselmann},\ and\ \citenamefont {Beling}}]{Gao:23}%
  \BibitemOpen
  \bibfield  {author} {\bibinfo {author} {\bibfnamefont {J.}~\bibnamefont {Gao}}, \bibinfo {author} {\bibfnamefont {T.~C.}\ \bibnamefont {Tzu}}, \bibinfo {author} {\bibfnamefont {T.}~\bibnamefont {Fatema}}, \bibinfo {author} {\bibfnamefont {X.}~\bibnamefont {Guo}}, \bibinfo {author} {\bibfnamefont {Q.}~\bibnamefont {Yu}}, \bibinfo {author} {\bibfnamefont {G.}~\bibnamefont {Navickaite}}, \bibinfo {author} {\bibfnamefont {M.}~\bibnamefont {Zervas}}, \bibinfo {author} {\bibfnamefont {M.}~\bibnamefont {Geiselmann}},\ and\ \bibinfo {author} {\bibfnamefont {A.}~\bibnamefont {Beling}},\ }\bibfield  {title} {\bibinfo {title} {Heterogeneous balanced photodetector on silicon nitride with 30 ghz bandwidth and 26 db common mode rejection ratio},\ }in\ \href {https://doi.org/10.1364/OFC.2023.W2B.2} {\emph {\bibinfo {booktitle} {Optical Fiber Communication Conference (OFC) 2023}}}\ (\bibinfo  {publisher} {Optica Publishing Group},\ \bibinfo {year} {2023})\ p.\ \bibinfo {pages} {W2B.2}\BibitemShut {NoStop}%
\bibitem [{\citenamefont {Costanzo}\ \emph {et~al.}(2021)\citenamefont {Costanzo}, \citenamefont {Gao}, \citenamefont {Shen}, \citenamefont {Yu}, \citenamefont {Alabdulwahab}, \citenamefont {Beling},\ and\ \citenamefont {Bowers}}]{Costanzo:21}%
  \BibitemOpen
  \bibfield  {author} {\bibinfo {author} {\bibfnamefont {R.}~\bibnamefont {Costanzo}}, \bibinfo {author} {\bibfnamefont {J.}~\bibnamefont {Gao}}, \bibinfo {author} {\bibfnamefont {X.}~\bibnamefont {Shen}}, \bibinfo {author} {\bibfnamefont {Q.}~\bibnamefont {Yu}}, \bibinfo {author} {\bibfnamefont {A.}~\bibnamefont {Alabdulwahab}}, \bibinfo {author} {\bibfnamefont {A.}~\bibnamefont {Beling}},\ and\ \bibinfo {author} {\bibfnamefont {S.~M.}\ \bibnamefont {Bowers}},\ }\bibfield  {title} {\bibinfo {title} {Low-noise balanced photoreceiver with inp-on-si photodiodes and sige bicmos transimpedance amplifier},\ }\href {https://opg.optica.org/jlt/abstract.cfm?URI=jlt-39-14-4837} {\bibfield  {journal} {\bibinfo  {journal} {J. Lightwave Technol.}\ }\textbf {\bibinfo {volume} {39}},\ \bibinfo {pages} {4837} (\bibinfo {year} {2021})}\BibitemShut {NoStop}%
\bibitem [{\citenamefont {Bruynsteen}\ \emph {et~al.}(2021)\citenamefont {Bruynsteen}, \citenamefont {Vanhoecke}, \citenamefont {Bauwelinck},\ and\ \citenamefont {Yin}}]{Bruynsteen:21}%
  \BibitemOpen
  \bibfield  {author} {\bibinfo {author} {\bibfnamefont {C.}~\bibnamefont {Bruynsteen}}, \bibinfo {author} {\bibfnamefont {M.}~\bibnamefont {Vanhoecke}}, \bibinfo {author} {\bibfnamefont {J.}~\bibnamefont {Bauwelinck}},\ and\ \bibinfo {author} {\bibfnamefont {X.}~\bibnamefont {Yin}},\ }\bibfield  {title} {\bibinfo {title} {Integrated balanced homodyne photonic--electronic detector for beyond 20 ghz shot-noise-limited measurements},\ }\href {https://doi.org/10.1364/OPTICA.420973} {\bibfield  {journal} {\bibinfo  {journal} {Optica}\ }\textbf {\bibinfo {volume} {8}},\ \bibinfo {pages} {1146} (\bibinfo {year} {2021})}\BibitemShut {NoStop}%
\bibitem [{\citenamefont {Raffaelli}\ \emph {et~al.}(2018)\citenamefont {Raffaelli}, \citenamefont {Ferranti}, \citenamefont {Mahler}, \citenamefont {Sibson}, \citenamefont {Kennard}, \citenamefont {Santamato}, \citenamefont {Sinclair}, \citenamefont {Bonneau}, \citenamefont {Thompson},\ and\ \citenamefont {Matthews}}]{Raffaelli:18}%
  \BibitemOpen
  \bibfield  {author} {\bibinfo {author} {\bibfnamefont {F.}~\bibnamefont {Raffaelli}}, \bibinfo {author} {\bibfnamefont {G.}~\bibnamefont {Ferranti}}, \bibinfo {author} {\bibfnamefont {D.~H.}\ \bibnamefont {Mahler}}, \bibinfo {author} {\bibfnamefont {P.}~\bibnamefont {Sibson}}, \bibinfo {author} {\bibfnamefont {J.~E.}\ \bibnamefont {Kennard}}, \bibinfo {author} {\bibfnamefont {A.}~\bibnamefont {Santamato}}, \bibinfo {author} {\bibfnamefont {G.}~\bibnamefont {Sinclair}}, \bibinfo {author} {\bibfnamefont {D.}~\bibnamefont {Bonneau}}, \bibinfo {author} {\bibfnamefont {M.~G.}\ \bibnamefont {Thompson}},\ and\ \bibinfo {author} {\bibfnamefont {J.~C.~F.}\ \bibnamefont {Matthews}},\ }\bibfield  {title} {\bibinfo {title} {A homodyne detector integrated onto a photonic chip for measuring quantum states and generating random numbers},\ }\href {https://doi.org/10.1088/2058-9565/aaa38f} {\bibfield  {journal} {\bibinfo  {journal} {Quantum Science and Technology}\ }\textbf {\bibinfo {volume} {3}},\ \bibinfo {pages}
  {025003} (\bibinfo {year} {2018})}\BibitemShut {NoStop}%
\bibitem [{\citenamefont {Tasker}\ \emph {et~al.}(2024)\citenamefont {Tasker}, \citenamefont {Frazer}, \citenamefont {Ferranti},\ and\ \citenamefont {Matthews}}]{Tasker:24}%
  \BibitemOpen
  \bibfield  {author} {\bibinfo {author} {\bibfnamefont {J.~F.}\ \bibnamefont {Tasker}}, \bibinfo {author} {\bibfnamefont {J.}~\bibnamefont {Frazer}}, \bibinfo {author} {\bibfnamefont {G.}~\bibnamefont {Ferranti}},\ and\ \bibinfo {author} {\bibfnamefont {J.~C.~F.}\ \bibnamefont {Matthews}},\ }\bibfield  {title} {\bibinfo {title} {A bi-cmos electronic photonic integrated circuit quantum light detector},\ }\href {https://doi.org/10.1126/sciadv.adk6890} {\bibfield  {journal} {\bibinfo  {journal} {Science Advances}\ }\textbf {\bibinfo {volume} {10}},\ \bibinfo {pages} {eadk6890} (\bibinfo {year} {2024})}\BibitemShut {NoStop}%
\bibitem [{\citenamefont {Tasker}\ \emph {et~al.}(2021)\citenamefont {Tasker}, \citenamefont {Frazer}, \citenamefont {Ferranti} \emph {et~al.}}]{Tasker:21}%
  \BibitemOpen
  \bibfield  {author} {\bibinfo {author} {\bibfnamefont {J.~F.}\ \bibnamefont {Tasker}}, \bibinfo {author} {\bibfnamefont {J.}~\bibnamefont {Frazer}}, \bibinfo {author} {\bibfnamefont {G.}~\bibnamefont {Ferranti}}, \emph {et~al.},\ }\bibfield  {title} {\bibinfo {title} {Silicon photonics interfaced with integrated electronics for 9 ghz measurement of squeezed light},\ }\href {https://doi.org/10.1038/s41566-020-00715-5} {\bibfield  {journal} {\bibinfo  {journal} {Nature Photonics}\ }\textbf {\bibinfo {volume} {15}},\ \bibinfo {pages} {11} (\bibinfo {year} {2021})}\BibitemShut {NoStop}%
\bibitem [{\citenamefont {Gurses}\ \emph {et~al.}(2024)\citenamefont {Gurses}, \citenamefont {Sarkar}, \citenamefont {Davis},\ and\ \citenamefont {Hajimiri}}]{Gurses:24}%
  \BibitemOpen
  \bibfield  {author} {\bibinfo {author} {\bibfnamefont {V.}~\bibnamefont {Gurses}}, \bibinfo {author} {\bibfnamefont {D.}~\bibnamefont {Sarkar}}, \bibinfo {author} {\bibfnamefont {S.}~\bibnamefont {Davis}},\ and\ \bibinfo {author} {\bibfnamefont {A.}~\bibnamefont {Hajimiri}},\ }\bibfield  {title} {\bibinfo {title} {An integrated photonic-electronic quantum coherent receiver for sub-shot-noise-limited optical links},\ }in\ \href {https://doi.org/10.1364/OFC.2024.Tu2C.1} {\emph {\bibinfo {booktitle} {Optical Fiber Communication Conference (OFC) 2024}}}\ (\bibinfo  {publisher} {Optica Publishing Group},\ \bibinfo {year} {2024})\ p.\ \bibinfo {pages} {Tu2C.1}\BibitemShut {NoStop}%
\bibitem [{\citenamefont {Lustig}\ \emph {et~al.}(2024)\citenamefont {Lustig}, \citenamefont {Guidry}, \citenamefont {Lukin}, \citenamefont {Fan},\ and\ \citenamefont {Vuckovic}}]{lustig_emerging_2024}%
  \BibitemOpen
  \bibfield  {author} {\bibinfo {author} {\bibfnamefont {E.}~\bibnamefont {Lustig}}, \bibinfo {author} {\bibfnamefont {M.~A.}\ \bibnamefont {Guidry}}, \bibinfo {author} {\bibfnamefont {D.~M.}\ \bibnamefont {Lukin}}, \bibinfo {author} {\bibfnamefont {S.}~\bibnamefont {Fan}},\ and\ \bibinfo {author} {\bibfnamefont {J.}~\bibnamefont {Vuckovic}},\ }\bibfield  {title} {\bibinfo {title} {Emerging {Quadrature} {Lattices} of {Kerr} {Combs}},\ }\bibfield  {journal} {\bibinfo  {journal} {arXiv:2407.13049}\ }\href {https://doi.org/10.48550/arXiv.2407.13049} {10.48550/arXiv.2407.13049} (\bibinfo {year} {2024})\BibitemShut {NoStop}%
\bibitem [{\citenamefont {Pu}\ \emph {et~al.}(2018)\citenamefont {Pu}, \citenamefont {Hu}, \citenamefont {Ottaviano}, \citenamefont {Semenova}, \citenamefont {Vukovic}, \citenamefont {Oxenløwe},\ and\ \citenamefont {Yvind}}]{pu_ultra-efficient_2018}%
  \BibitemOpen
  \bibfield  {author} {\bibinfo {author} {\bibfnamefont {M.}~\bibnamefont {Pu}}, \bibinfo {author} {\bibfnamefont {H.}~\bibnamefont {Hu}}, \bibinfo {author} {\bibfnamefont {L.}~\bibnamefont {Ottaviano}}, \bibinfo {author} {\bibfnamefont {E.}~\bibnamefont {Semenova}}, \bibinfo {author} {\bibfnamefont {D.}~\bibnamefont {Vukovic}}, \bibinfo {author} {\bibfnamefont {L.~K.}\ \bibnamefont {Oxenløwe}},\ and\ \bibinfo {author} {\bibfnamefont {K.}~\bibnamefont {Yvind}},\ }\bibfield  {title} {\bibinfo {title} {Ultra-{Efficient} and {Broadband} {Nonlinear} {AlGaAs}-on-{Insulator} {Chip} for {Low}-{Power} {Optical} {Signal} {Processing}},\ }\href {https://doi.org/10.1002/lpor.201800111} {\bibfield  {journal} {\bibinfo  {journal} {Laser \& Photonics Reviews}\ }\textbf {\bibinfo {volume} {12}},\ \bibinfo {pages} {1800111} (\bibinfo {year} {2018})}\BibitemShut {NoStop}%
\bibitem [{\citenamefont {Xie}\ \emph {et~al.}(2020)\citenamefont {Xie}, \citenamefont {Chang}, \citenamefont {Shu}, \citenamefont {Norman}, \citenamefont {Peters}, \citenamefont {Wang},\ and\ \citenamefont {Bowers}}]{xie_ultrahigh-q_2020}%
  \BibitemOpen
  \bibfield  {author} {\bibinfo {author} {\bibfnamefont {W.}~\bibnamefont {Xie}}, \bibinfo {author} {\bibfnamefont {L.}~\bibnamefont {Chang}}, \bibinfo {author} {\bibfnamefont {H.}~\bibnamefont {Shu}}, \bibinfo {author} {\bibfnamefont {J.~C.}\ \bibnamefont {Norman}}, \bibinfo {author} {\bibfnamefont {J.~D.}\ \bibnamefont {Peters}}, \bibinfo {author} {\bibfnamefont {X.}~\bibnamefont {Wang}},\ and\ \bibinfo {author} {\bibfnamefont {J.~E.}\ \bibnamefont {Bowers}},\ }\bibfield  {title} {\bibinfo {title} {Ultrahigh-{Q} {AlGaAs}-on-insulator microresonators for integrated nonlinear photonics},\ }\href {https://doi.org/10.1364/OE.405343} {\bibfield  {journal} {\bibinfo  {journal} {Opt. Express}\ }\textbf {\bibinfo {volume} {28}},\ \bibinfo {pages} {32894} (\bibinfo {year} {2020})}\BibitemShut {NoStop}%
\bibitem [{\citenamefont {Guidry}\ \emph {et~al.}(2023)\citenamefont {Guidry}, \citenamefont {Lukin}, \citenamefont {Yang},\ and\ \citenamefont {Vučković}}]{Guidry:23}%
  \BibitemOpen
  \bibfield  {author} {\bibinfo {author} {\bibfnamefont {M.~A.}\ \bibnamefont {Guidry}}, \bibinfo {author} {\bibfnamefont {D.~M.}\ \bibnamefont {Lukin}}, \bibinfo {author} {\bibfnamefont {K.~Y.}\ \bibnamefont {Yang}},\ and\ \bibinfo {author} {\bibfnamefont {J.}~\bibnamefont {Vučković}},\ }\bibfield  {title} {\bibinfo {title} {Multimode squeezing in soliton crystal microcombs},\ }\href@noop {} {\bibfield  {journal} {\bibinfo  {journal} {Optica}\ }\textbf {\bibinfo {volume} {10}},\ \bibinfo {pages} {694} (\bibinfo {year} {2023})}\BibitemShut {NoStop}%
\bibitem [{\citenamefont {Pontula}\ \emph {et~al.}(2024)\citenamefont {Pontula}, \citenamefont {Salamin}, \citenamefont {Roques-Carmes},\ and\ \citenamefont {Soljacic}}]{pontula_2024_multimode}%
  \BibitemOpen
  \bibfield  {author} {\bibinfo {author} {\bibfnamefont {S.}~\bibnamefont {Pontula}}, \bibinfo {author} {\bibfnamefont {Y.}~\bibnamefont {Salamin}}, \bibinfo {author} {\bibfnamefont {C.}~\bibnamefont {Roques-Carmes}},\ and\ \bibinfo {author} {\bibfnamefont {M.}~\bibnamefont {Soljacic}},\ }\bibfield  {title} {\bibinfo {title} {Multimode amplitude squeezing through cascaded nonlinear optical processes},\ }\href {https://arxiv.org/abs/2405.05201} {\bibfield  {journal} {\bibinfo  {journal} {arXiv:2405.05201}\ } (\bibinfo {year} {2024})}\BibitemShut {NoStop}%
\bibitem [{\citenamefont {Moody}\ \emph {et~al.}(2022)\citenamefont {Moody}, \citenamefont {Sorger}, \citenamefont {Blumenthal}, \citenamefont {Juodawlkis}, \citenamefont {Loh}, \citenamefont {Sorace-Agaskar}, \citenamefont {Jones}, \citenamefont {Balram}, \citenamefont {Matthews}, \citenamefont {Laing}, \citenamefont {Davanco}, \citenamefont {Chang}, \citenamefont {Bowers}, \citenamefont {Quack}, \citenamefont {Galland}, \citenamefont {Aharonovich}, \citenamefont {Wolff}, \citenamefont {Schuck}, \citenamefont {Sinclair}, \citenamefont {Lončar}, \citenamefont {Komljenovic}, \citenamefont {Weld}, \citenamefont {Mookherjea}, \citenamefont {Buckley}, \citenamefont {Radulaski}, \citenamefont {Reitzenstein}, \citenamefont {Pingault}, \citenamefont {Machielse}, \citenamefont {Mukhopadhyay}, \citenamefont {Akimov}, \citenamefont {Zheltikov}, \citenamefont {Agarwal}, \citenamefont {Srinivasan}, \citenamefont {Lu}, \citenamefont {Tang}, \citenamefont {Jiang}, \citenamefont {McKenna}, \citenamefont {Safavi-Naeini},
  \citenamefont {Steinhauer}, \citenamefont {Elshaari}, \citenamefont {Zwiller}, \citenamefont {Davids}, \citenamefont {Martinez}, \citenamefont {Gehl}, \citenamefont {Chiaverini}, \citenamefont {Mehta}, \citenamefont {Romero}, \citenamefont {Lingaraju}, \citenamefont {Weiner}, \citenamefont {Peace}, \citenamefont {Cernansky}, \citenamefont {Lobino}, \citenamefont {Diamanti}, \citenamefont {Vidarte},\ and\ \citenamefont {Camacho}}]{moody_2022_2022}%
  \BibitemOpen
  \bibfield  {author} {\bibinfo {author} {\bibfnamefont {G.}~\bibnamefont {Moody}}, \bibinfo {author} {\bibfnamefont {V.~J.}\ \bibnamefont {Sorger}}, \bibinfo {author} {\bibfnamefont {D.~J.}\ \bibnamefont {Blumenthal}}, \bibinfo {author} {\bibfnamefont {P.~W.}\ \bibnamefont {Juodawlkis}}, \bibinfo {author} {\bibfnamefont {W.}~\bibnamefont {Loh}}, \bibinfo {author} {\bibfnamefont {C.}~\bibnamefont {Sorace-Agaskar}}, \bibinfo {author} {\bibfnamefont {A.~E.}\ \bibnamefont {Jones}}, \bibinfo {author} {\bibfnamefont {K.~C.}\ \bibnamefont {Balram}}, \bibinfo {author} {\bibfnamefont {J.~C.~F.}\ \bibnamefont {Matthews}}, \bibinfo {author} {\bibfnamefont {A.}~\bibnamefont {Laing}}, \bibinfo {author} {\bibfnamefont {M.}~\bibnamefont {Davanco}}, \bibinfo {author} {\bibfnamefont {L.}~\bibnamefont {Chang}}, \bibinfo {author} {\bibfnamefont {J.~E.}\ \bibnamefont {Bowers}}, \bibinfo {author} {\bibfnamefont {N.}~\bibnamefont {Quack}}, \bibinfo {author} {\bibfnamefont {C.}~\bibnamefont {Galland}}, \bibinfo {author}
  {\bibfnamefont {I.}~\bibnamefont {Aharonovich}}, \bibinfo {author} {\bibfnamefont {M.~A.}\ \bibnamefont {Wolff}}, \bibinfo {author} {\bibfnamefont {C.}~\bibnamefont {Schuck}}, \bibinfo {author} {\bibfnamefont {N.}~\bibnamefont {Sinclair}}, \bibinfo {author} {\bibfnamefont {M.}~\bibnamefont {Lončar}}, \bibinfo {author} {\bibfnamefont {T.}~\bibnamefont {Komljenovic}}, \bibinfo {author} {\bibfnamefont {D.}~\bibnamefont {Weld}}, \bibinfo {author} {\bibfnamefont {S.}~\bibnamefont {Mookherjea}}, \bibinfo {author} {\bibfnamefont {S.}~\bibnamefont {Buckley}}, \bibinfo {author} {\bibfnamefont {M.}~\bibnamefont {Radulaski}}, \bibinfo {author} {\bibfnamefont {S.}~\bibnamefont {Reitzenstein}}, \bibinfo {author} {\bibfnamefont {B.}~\bibnamefont {Pingault}}, \bibinfo {author} {\bibfnamefont {B.}~\bibnamefont {Machielse}}, \bibinfo {author} {\bibfnamefont {D.}~\bibnamefont {Mukhopadhyay}}, \bibinfo {author} {\bibfnamefont {A.}~\bibnamefont {Akimov}}, \bibinfo {author} {\bibfnamefont {A.}~\bibnamefont {Zheltikov}},
  \bibinfo {author} {\bibfnamefont {G.~S.}\ \bibnamefont {Agarwal}}, \bibinfo {author} {\bibfnamefont {K.}~\bibnamefont {Srinivasan}}, \bibinfo {author} {\bibfnamefont {J.}~\bibnamefont {Lu}}, \bibinfo {author} {\bibfnamefont {H.~X.}\ \bibnamefont {Tang}}, \bibinfo {author} {\bibfnamefont {W.}~\bibnamefont {Jiang}}, \bibinfo {author} {\bibfnamefont {T.~P.}\ \bibnamefont {McKenna}}, \bibinfo {author} {\bibfnamefont {A.~H.}\ \bibnamefont {Safavi-Naeini}}, \bibinfo {author} {\bibfnamefont {S.}~\bibnamefont {Steinhauer}}, \bibinfo {author} {\bibfnamefont {A.~W.}\ \bibnamefont {Elshaari}}, \bibinfo {author} {\bibfnamefont {V.}~\bibnamefont {Zwiller}}, \bibinfo {author} {\bibfnamefont {P.~S.}\ \bibnamefont {Davids}}, \bibinfo {author} {\bibfnamefont {N.}~\bibnamefont {Martinez}}, \bibinfo {author} {\bibfnamefont {M.}~\bibnamefont {Gehl}}, \bibinfo {author} {\bibfnamefont {J.}~\bibnamefont {Chiaverini}}, \bibinfo {author} {\bibfnamefont {K.~K.}\ \bibnamefont {Mehta}}, \bibinfo {author} {\bibfnamefont
  {J.}~\bibnamefont {Romero}}, \bibinfo {author} {\bibfnamefont {N.~B.}\ \bibnamefont {Lingaraju}}, \bibinfo {author} {\bibfnamefont {A.~M.}\ \bibnamefont {Weiner}}, \bibinfo {author} {\bibfnamefont {D.}~\bibnamefont {Peace}}, \bibinfo {author} {\bibfnamefont {R.}~\bibnamefont {Cernansky}}, \bibinfo {author} {\bibfnamefont {M.}~\bibnamefont {Lobino}}, \bibinfo {author} {\bibfnamefont {E.}~\bibnamefont {Diamanti}}, \bibinfo {author} {\bibfnamefont {L.~T.}\ \bibnamefont {Vidarte}},\ and\ \bibinfo {author} {\bibfnamefont {R.~M.}\ \bibnamefont {Camacho}},\ }\bibfield  {title} {\bibinfo {title} {2022 {Roadmap} on integrated quantum photonics},\ }\href {https://doi.org/10.1088/2515-7647/ac1ef4} {\bibfield  {journal} {\bibinfo  {journal} {J. Phys. Photonics}\ }\textbf {\bibinfo {volume} {4}},\ \bibinfo {pages} {012501} (\bibinfo {year} {2022})}\BibitemShut {NoStop}%
\end{thebibliography}%

\end{document}